\documentclass[aps,prl,showpacs,twocolumn]{revtex4-1}

\usepackage{dcolumn}
\usepackage{graphicx}
\usepackage{amssymb,amsmath}
\usepackage{natbib}
\usepackage{subfigure}
\usepackage{epstopdf}
\usepackage{placeins}
\usepackage{color}

\newcommand{\tpr}{\text{Pr}}
\newcommand{\tra}{\text{Ra}}
\newcommand{\tnu}{\text{Nu}}

\begin{document}

\title{The effect of velocity boundary conditions on the heat transfer and flow topology in two-dimensional Rayleigh-B\'enard convection}
\author{Erwin P. van der Poel$^{1}$, Rodolfo Ostilla M\'onico$^{1}$, Roberto Verzicco$^{1,2}$ and Detlef Lohse$^{1}$}
\affiliation{
$^1$Department of Physics, Mesa+ Institute,  and J.\ M.\ Burgers Centre for Fluid Dynamics, University of Twente, 7500 AE Enschede, The Netherlands \\
$^2$Dipartimento di Ingegneria Industriale, University of Rome ``Tor Vergata'', Via del Politecnico 1, Roma 00133, Italy}

\date{\today}

\begin{abstract}
The effect of various velocity boundary condition is studied in two-dimensional Rayleigh-B\'enard convection. Combinations of no-slip, stress-free and periodic boundary conditions are used on both the sidewalls and the horizontal plates. For the studied Rayleigh numbers $\tra$ between $10^8$ and $10^{11}$ the heat transport is lower for $\Gamma = 0.33$ than for $\Gamma = 1$ in case of no-slip sidewalls. This is surprisingly opposite for stress-free sidewalls, where the heat transport increases for lower aspect-ratio. In wider cells the aspect-ratio dependence is observed to disappear for $\tra \ge 10^{10}$. Two distinct flow types with very different dynamics can be seen, mostly dependent on the plate velocity boundary condition, namely roll-like flow and horizontal zonal flow, which have a substantial effect on the dynamics and heat transport in the system. The predominantly horizontal zonal flow suppresses heat flux and is observed for stress-free and asymmetric plates. Low aspect-ratio periodic sidewall simulations with a no-slip boundary condition on the plates also exhibit zonal flow. In all the other cases, the flow is roll-like. In two-dimensional Rayleigh-B\'enard convection, the velocity boundary conditions thus have large implications on both roll-like and zonal flow that have to be taken into consideration before the boundary conditions are imposed.
\end{abstract}

\maketitle

\section{Introduction}

\begin{figure*}
\centering
\subfigure{\includegraphics[height=0.55\textwidth]{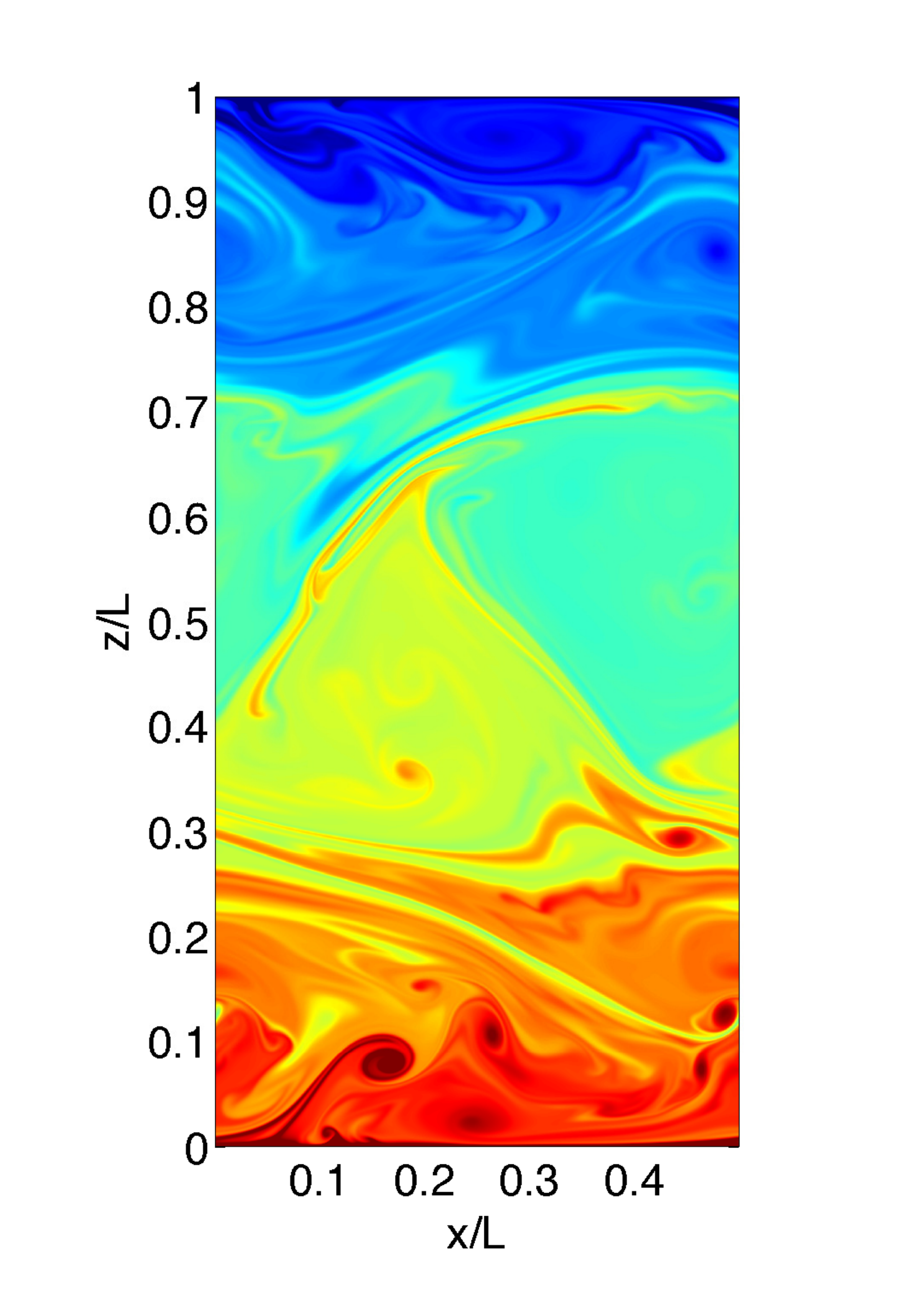}}
\subfigure{\includegraphics[height=0.55\textwidth]{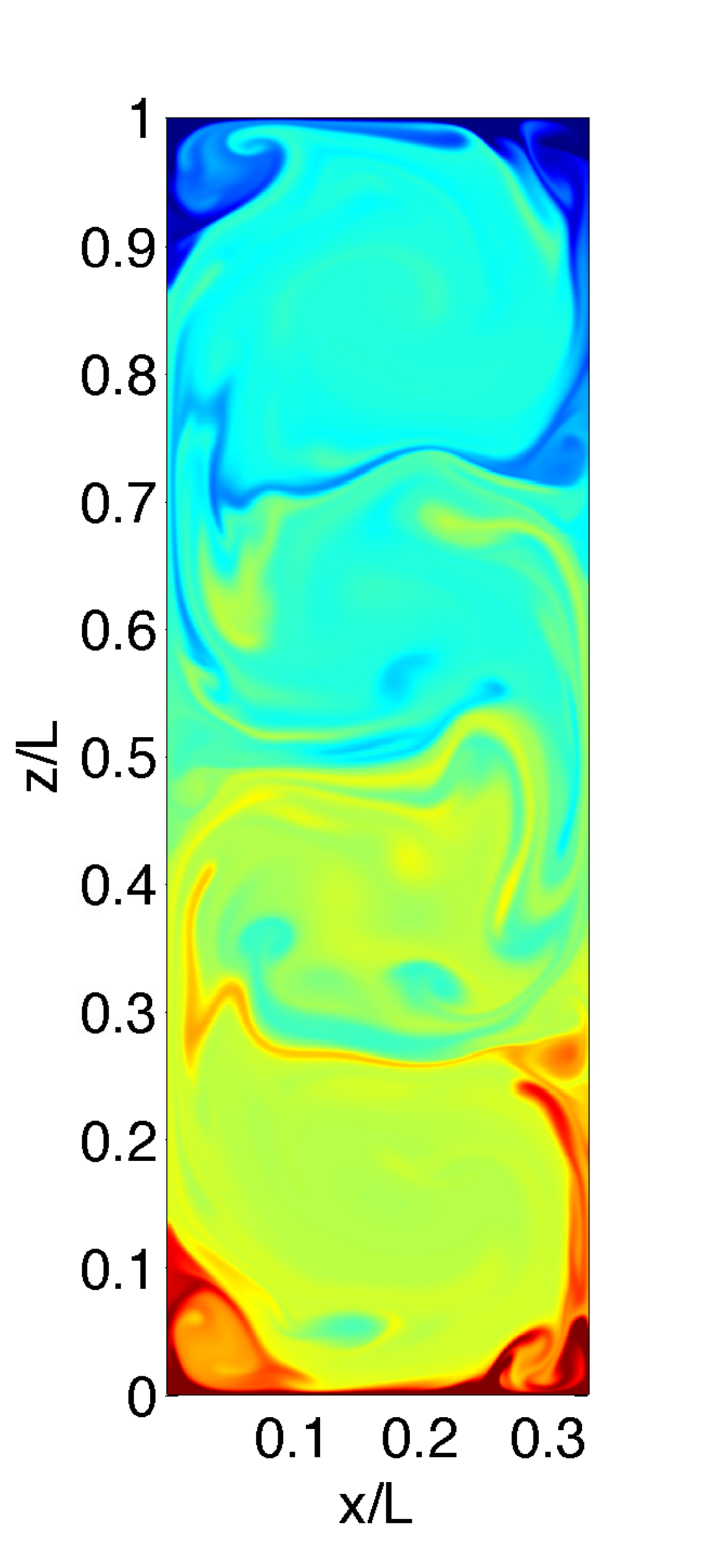}}
\subfigure{\includegraphics[height=0.55\textwidth]{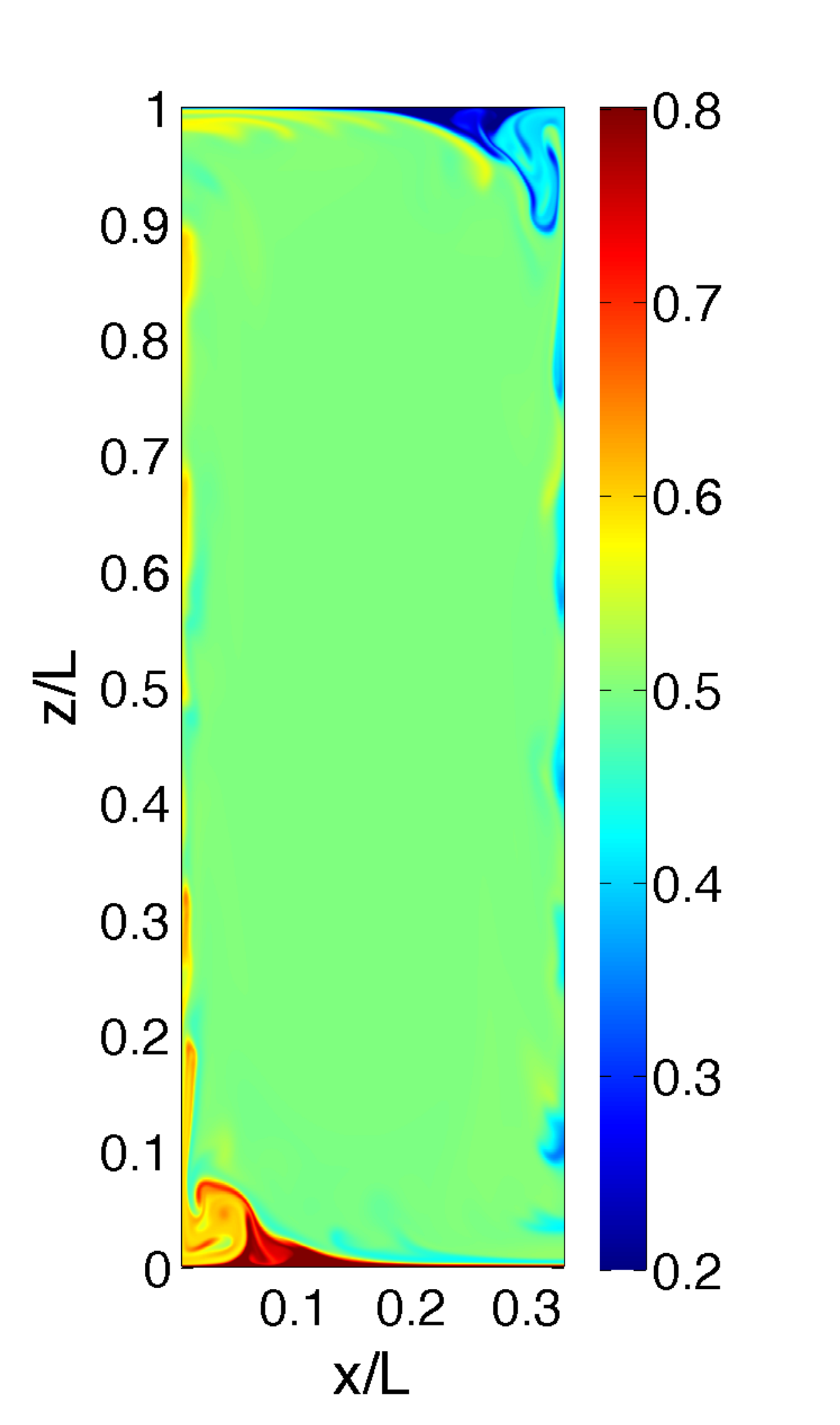}}
\caption{(Color online) Temperature field snapshots for different simulations. Red and blue indicate hot and cold fluid respectively. The colour varies between $\theta = 0.2$ and $\theta = 0.8$. $\tra = 10^{11}$ and $\Gamma = 1/2$ for lateral periodicity, showing zonal flow (left panel), $\tra = 10^{10}$ and $\Gamma = 0.33$ for no-slip sidewalls, showing roll structures (center panel), and $\tra = 10^{10}$ and $\Gamma = 0.33$ for stress-free sidewalls (right panel). The plates are no-slip in all cases. Movies can be found in the supplementary material.}
\label{fig:PD3}
\end{figure*}

Natural convection is commonly modelled by the Rayleigh-B\'enard (RB) system \cite{ahl09,chi12,sig94,loh10} to reduce the complexity of the problem while preserving the most interesting features. This is a system in which a fluid is heated from below and cooled from above; gravity is acting in the plane normal to the energy-injecting or energy-extractinglower and upper plates. The temperature induced density differences result in a (temperature dependent) vertical forcing. Under the Boussinesq approximation, in which the fluid properties are assumed to be independent of temperature except for the buoyancy term, the governing equations in dimensionless form are:
\begin{eqnarray}
\label{eq:momentum}
\partial_tu_i + u_j\partial_ju_i = -\partial_ip + \sqrt{\frac{\tpr}{\tra}}\partial^2_ju_i + \theta\hat{e}_3, \\
\label{eq:thermal}
\partial_t\theta + u_j\partial_j\theta = \frac{1}{\sqrt{\tra\tpr}}\partial^2_j\theta,
\end{eqnarray}
where $\hat{e}_3$ is the unit vector anti-parallel to gravity, $u_i$ is the velocity normalized by the freefall velocity $\sqrt{g\beta\Delta L}$, t is the time normalized by the freefall time $\sqrt{L / (g\beta\Delta)}$ and $\theta$ is the temperature normalized by $\Delta$, the temperature difference between top and bottom plate, and shifted such that $\theta$ is in the range $0 \leq \theta \leq 1$. The control parameters of the system are the non-dimensional temperature difference, i.e. the Rayleigh number $\tra = g\beta\Delta L^3 / (\nu\kappa) $, the Prandtl number of the fluid $\tpr = \nu/\kappa$ and the aspect-ratio $\Gamma = D/L$, where $L$ is the height of the sample and $D$ its width, $g$ the gravitational acceleration, $\beta$ is the thermal expansion coefficient and $\nu$ and $\kappa$ the kinematic viscosity and the thermal diffusivity, respectively. 

RB convection is used as a simple model for all types of natural convection, regardless of geometry. For 2D RB, at a moderate $\tra = 10^8$, the heat flux is expected to be independent of the aspect-ratio, and thus of the sidewall boundary conditions, beyond $\Gamma \gtrsim 26$ \cite{poe12}. This value of $\Gamma$ is expected to be lower for 3D RB \cite{bai10}. Low aspect-ratio applications with a no-slip BC on the lateral walls, such as cooling fins and domestic heating, have moderate Rayleigh numbers ($\tra < 10^{10}$) that are accessible with contemporary direct numerical simulations (DNS). However, Nature is dominated by flows with large aspect-ratios and inaccessible high Ra \cite{chi12}. 

In both experiments and numerics, the aspect-ratio $\Gamma$ is commonly reduced to decrease the volume of the setup, as Ra depends only on the height of the system. In simulations, reducing the aspect-ratio for fixed Ra decreases the computational complexity by $\mathcal{O}(\Gamma^2)$ in three-dimensions, while in experiments it means a smaller setup volume. In both cases it makes higher Rayleigh numbers more accesible. Despite the additional effort required for large aspect-ratio, research in this direction has been done in simulations \cite{bai10} and experiments \cite{fun05,pui13,hog13}. However, these endeavours were all limited in Ra due the large aspect-ratios used. In simulations one can approximate infinitely large $\Gamma$ by using periodic boundary conditions at the sidewalls. This leads to the question how large the simulated domain has to be to sufficiently approximate infinite $\Gamma$, as a truly infinite domain can in principle support any wavenumber while a periodic domain allows only for multiples of the fundamental one imposed by its size. This is addressed in this study in 2D RB, which is expected to require a larger domain than 3D and therefore is more demanding in terms of computational resources.

Different temperature boundary conditions at the plates were studied by \cite{joh09}, where no significant difference between constant flux and isothermal plates was found in the higher Ra regime. Therefore, in this manuscript we focus only on the \emph{velocity} boundary conditions. High Rayleigh number applications of RB usually do not have a no-slip lateral wall BC. Such lateral walls are unfortunately unavoidable in experiments due to flow confinement, but it is not required in numerics, although numerics tend to use no-slip sidewalls when replicating or complimenting experiments. Removing the no-slip BC on the lateral wall is actually beneficial in simulations, as it reduces computational requirements owing to the absence of the viscous BL and therefore no need to cluster grid points in the wall-normal direction. In any case, boundary conditions must be chosen in numerical studies and one has to chose the boundary best suited for the intended study, but also the computational demand and possible introduced artifacts have to be taken into account. 

\begin{figure*}
\centering
\subfigure{\includegraphics[width=0.89\textwidth]{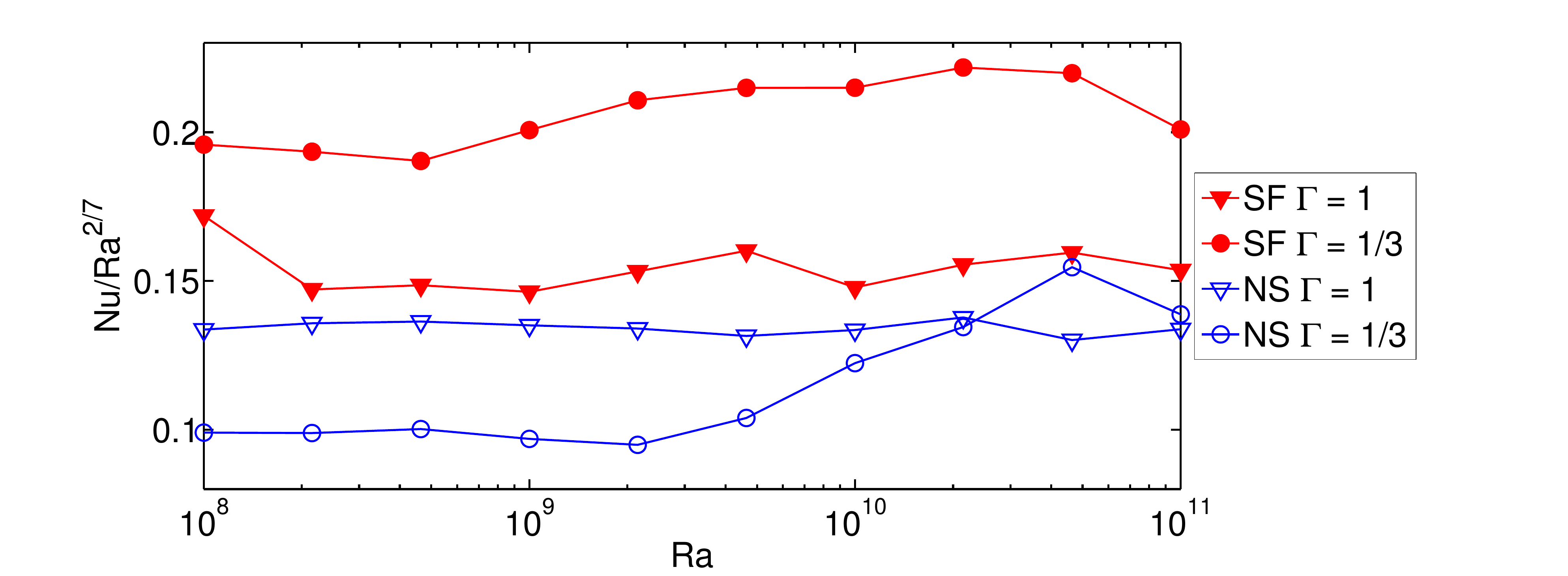}}
\caption{(Color online) Compensated $\text{Nu}/\tra^{2/7}$ as a function of Ra for stress-free and no-slip sidewalls and no-slip plates. The data is for aspect-ratios $\Gamma = 1$ and $\Gamma = 0.33$. As expected, the stress-free Nu is consistently higher than for no-slip for the same value of $\Gamma$. For stress-free boundary conditions, the heat transport is higher for $\Gamma = 0.33$ than for $\Gamma = 1$ for all of the evaluated Ra, while for no-slip this is only observed at high Ra.}
\label{fig:FSNS}
\end{figure*}

Despite its perceived importance, the effect of the velocity boundary conditions on the lateral walls has not been systematically studied for RB even though it is expected to have a significant effect on the roll state, absolute heat flux and kinetic energy of the flow. In numerics, any expression which can be mathematically formulated can be used as a boundary condition, but here we focus on the three most physical options: no-slip (NS), stress-free (SF) and periodic (PD). For no-slip and stress-free sidewalls, an additional temperature BC has to be imposed which in our case is adiabatic. 

Many recent studies on RB convection focus on the so-called ``ultimate regime'' \cite{kra67,gro11,loh03,roc10,he12}, a regime of RB flow where the boundary layers are expected to become fully turbulent. It is expected that the scaling of integral quantities in this regime can be extrapolated to the very high Ra numbers present in geo- and astrophysical natural convection. In this manuscript, we address the question of how the lateral wall boundary conditions affect the flow and how we can extrapolate the results obtained to flows at arbitrary large driving, such as those in the ultimate regime. We find that, in essence, when one strives to model the high Ra applications of natural convection, the no-slip lateral wall BC and the low aspect-ratios of experiments and simulations may obscure phenomena which are seen in large aspect-ratio convection found in Nature. This research focuses on two-dimensional RB and it must be noted that three-dimensional RB might behave differently. In specific cases, qualitative arguments result in the expectation that three-dimensional RB does not show phenomena at all that are very strong in two-dimensional RB \cite{poe13}. 

One of these phenomena is zonal flow, which is a flow state with most of the kinetic energy stored in the horizontal motion \cite{kri81,gol14}. The topology consists of two shearing layers of fluid, one at the top and one at the bottom, that move in opposite horizontal directions. Zonal flow strongly suppresses heat flux, and correspondingly the heat flux as a function of time behaves very differently. It can either be bursting or be sustained. In the first case quiescent periods of negligible convective heat flux are interrupted by large bursts, while in the second case the heat flux is more sustained. Even in the sustained case, the heat flux is much lower than would be expected in the more conventional roll-like flow state with identical control parameters. Zonal flow is observed with stress-free, asymmetric and no-slip plates, though for no-slip plates, the size of the period in laterally periodic simulations has to be small. 

The organization of the manuscript is as follows: First, the boundary conditions are formulated and their consequences on the numerical algorithm are evaluated. Then, simulations dominated by roll-like dynamics are presented in both thin ($\Gamma < 1$) and wide ($\Gamma > 1$) cells. Here, we compare the heat flux in no-slip, stress-free and periodic sidewalls as function of Ra. We quantify the behaviour of the heat flux as function of $\Gamma$ for periodic boundary conditions in order to find the aspect-ratio for which periodicity does not introduce any artifacts. Finally, we study the effect of stress-free plates, both in an asymmetric setup with only one stress-free plate, and also with both plates having stress-free velocity boundary conditions. Here the heat flux as a function of time is dominated by large bursts with quiescent intervals. For all simulations a Prandtl number equal to unity is used, unless stated otherwise, to prevent the nesting of the kinetic into the thermal boundary layer and vice versa, and to reduce the parameter space to a manageable size.

\section{Numerical method} \label{sect:numerics}

The direct numerical simulations (DNS) were performed with a two-dimensional version of a second order finite difference scheme \cite{ver99,ver03}. The underlying numerical scheme has been used for simulations of three-dimensional RB \cite{ste10a,bai10}, Taylor-Couette \cite{ost13} and other flows \cite{lac03}. The two-dimensional version of the code was validated by comparison to a fourth order finite difference code \cite{sug09} that on its turn was compared with quasi two-dimensional experiments. Specifically, the $\tnu(\tra)$ data for $\Gamma = 0.33$ with no-slip boundary conditions can be compared one-to-one with the fourth order code and turn out to be equal within error \cite{poe12}. In addition, results for no-slip boundary conditions in a $\Gamma = 1$ cell at $\tpr = 1$ agree well with the fourth order compact scheme of \cite{hua13b}.

\begin{figure*}
\centering
\subfigure{\includegraphics[width=0.45\textwidth]{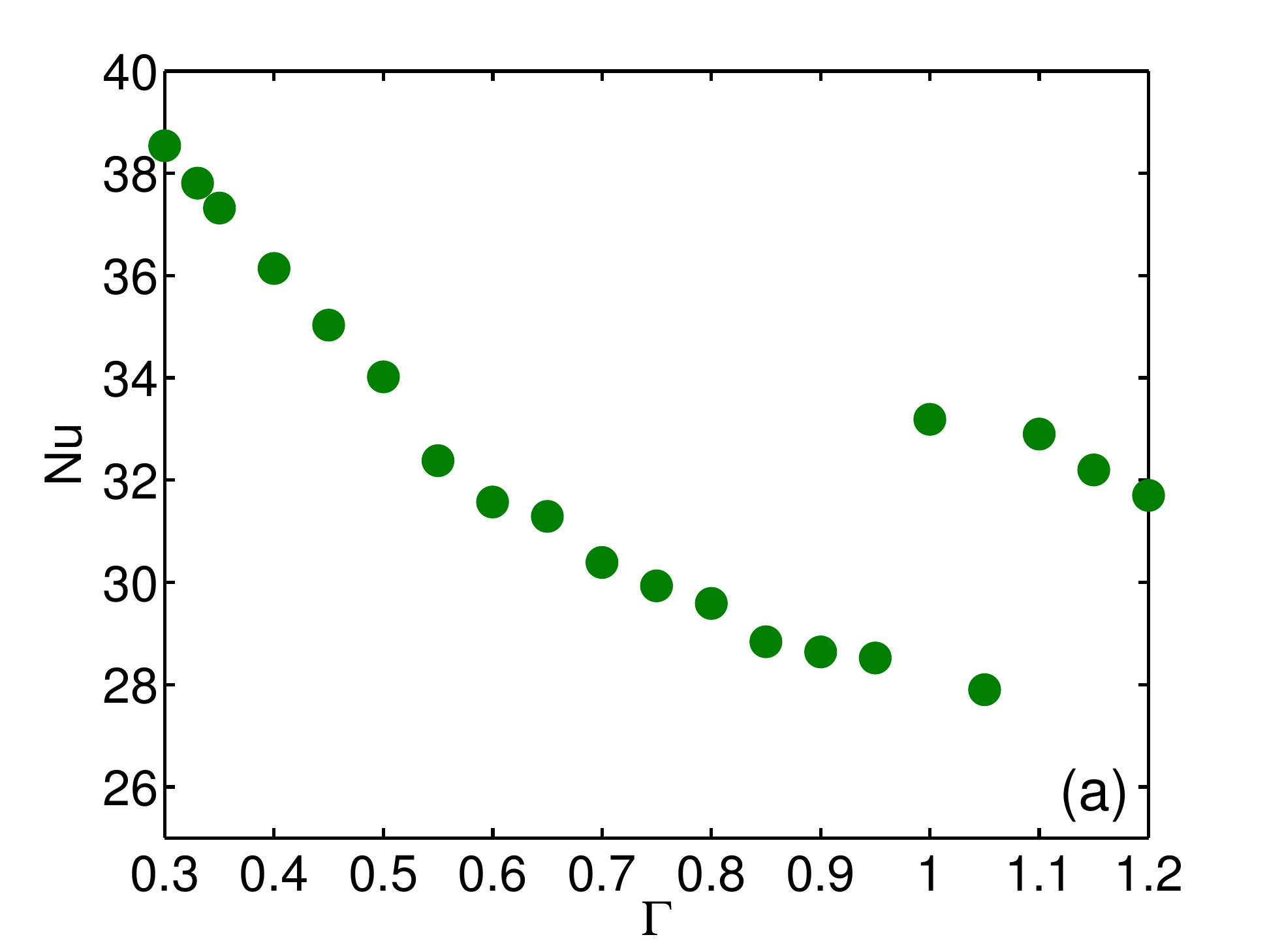}}
\subfigure{\includegraphics[width=0.45\textwidth]{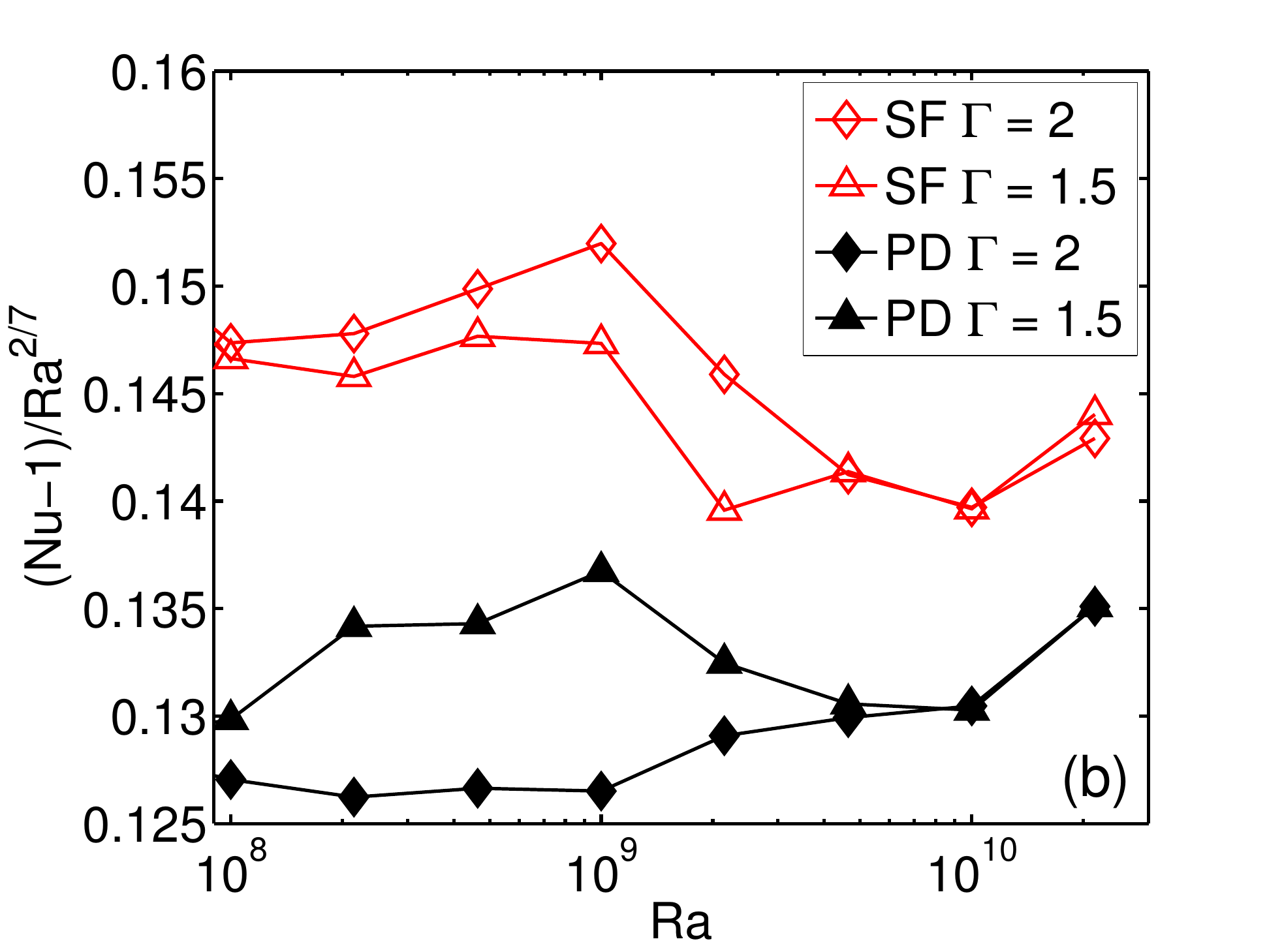}}
\caption{(Color online) a) Nu as a function of $\Gamma$ for stress-free sidewalls and no-slip plates. The data is for $\tra = 10^8$. At $\Gamma \approx 1$ there is a transition region between a roll-state with one roll and a roll state with two horizontally aligned rolls, which is reflected by hysteretic jumps in Nu. b) Compensated $(\tnu-1)/\tra^{2/7}$ as a function of Ra for stress-free and periodic sidewalls and no-slip plates. Data is depicted for $\Gamma = 1.5$ and $\Gamma = 2$. The effect of the aspect-ratio on Nu disappears for high Ra while the effect of the sidewall boundary condition remains.}
\label{fig:FS}
\end{figure*}

The Nusselt number Nu is calculated using
\begin{equation}
\tnu = \sqrt{\tra\tpr}\langle u_z \theta \rangle_{x,t} - \langle \partial_z \theta \rangle_{x,t},
\end{equation}
where $x$ and $z$ are the horizontal and vertical coordinate, respectively. Nu is a time average and one could denote it as $\langle \tnu \rangle_{t}$ and remove the time average from its definition, however Nu is used conform to RB literature. For the instanteneous Nu, the notation $\tnu(t)$ is used. Nu is calculated as both an area average of the convective term $\tnu = \sqrt{\tra\tpr}\langle u_z \rangle_{A,t} + 1$ and a line average on both the horizontal plates of the diffusive term $\tnu = -\langle \partial_z \theta |_{z=0,L} \rangle_{x,t}$. Note that the area average in two dimensions corresponds to a volume average in three dimensions. These Nu's are verified by calculating Nu through the kinetic and thermal dissipation rates \cite{shr90,sig94}. A discrepancy between any of the alltogether five different ways of calculating Nu indicates either insufficient statistical convergence or an underresolved simulation and so comparing these is a useful test. In addition, resolution checks, in which simulations are redundantly run using an increased resolution, are used to prove sufficient resolution for some cases. Additional details of the simulations can be found in the Appendix.

The boundary conditions chosen are either no-slip, i.e. $u_i = 0$ at the walls, stress-free walls, i.e. $u_i n_i = 0$, where $n_i$ is the vector normal to the wall, and $\partial_\tau u_i = 0$, where $\tau$ is the direction tangent to the wall, and lateral periodicity, i.e. $u_i(x = 0,z) = u_i(x = D,z)$. The temperature boundary condition on the plates are isothermal in all cases. For the no-slip and stress-free sidewalls, as additional temperature boundary condition we chose adiabaticity, i.e. $\partial_n \theta = 0$.

The use of different boundary conditions has a significant bearing on the algorithm to be chosen to solve the Boussinesq equations. The main difference is that for stress-free and periodic boundary conditions grids with uniformly spaced points can be used without loss of efficiency. In contrast, for no-slip, resolving the sharp gradients in the resulting boundary layers with a uniform grid-distribution will over-resolve the bulk. It is therefore more efficient to use a non-uniform grid to cluster points near the boundary layers. However, solving an equation on a non-uniform grid is generally more costly in terms of computational requirements when compared to solving the same equation on an equally sized uniform grid. This is by virtue of the uniform grid that allows for the use of a Fourier transform to solve the pressure equation to impose the incompressibility of the flow.

\section{Roll dominated convection}

\subsection{Thin cells}

For stress-free sidewalls, Nu for $\Gamma = 0.33$ is higher than for $\Gamma = 1$, which is in contrast with the no-slip case, where Nu for $\Gamma=1$ is higher than for $\Gamma=0.33$. It must be emphasized that this strong aspect-ratio dependence is a 2D effect not observed in 3D \cite{bai10}. Even for $\Gamma < 1$ it is observed in 3D experiments that there is no difference in Nu between $\Gamma = 1$ and $\Gamma = 1/2$ cells for identical Ra and Pr \cite{ahl12b,he12b}. This is despite the fact that both in 2D and 3D the amount of vertically stacked rolls changes with decreasing aspect-ratio \cite{poe13}. 

\begin{figure*}
\centering
\subfigure{\includegraphics[width=0.45\textwidth]{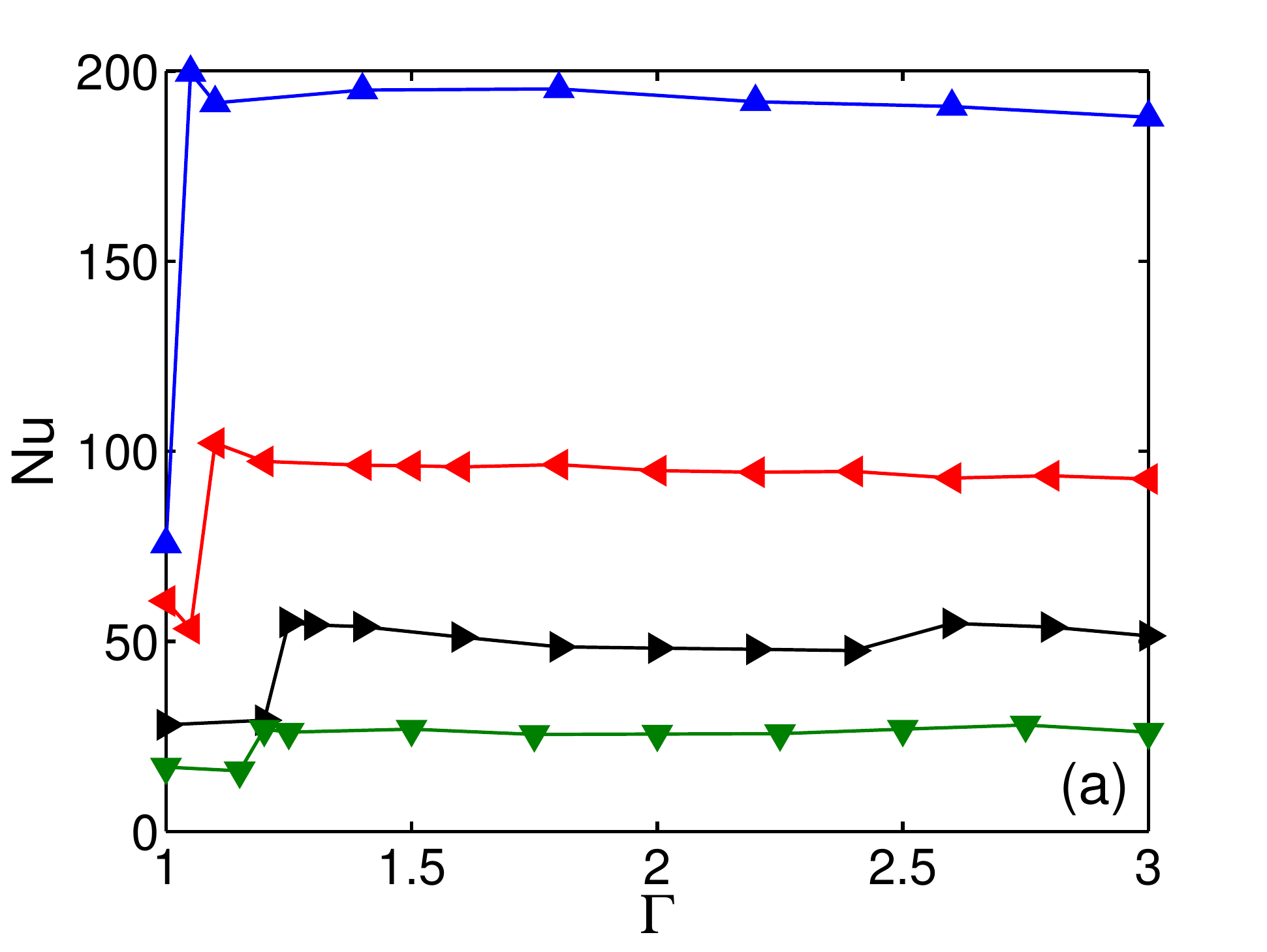}}
\subfigure{\includegraphics[width=0.45\textwidth]{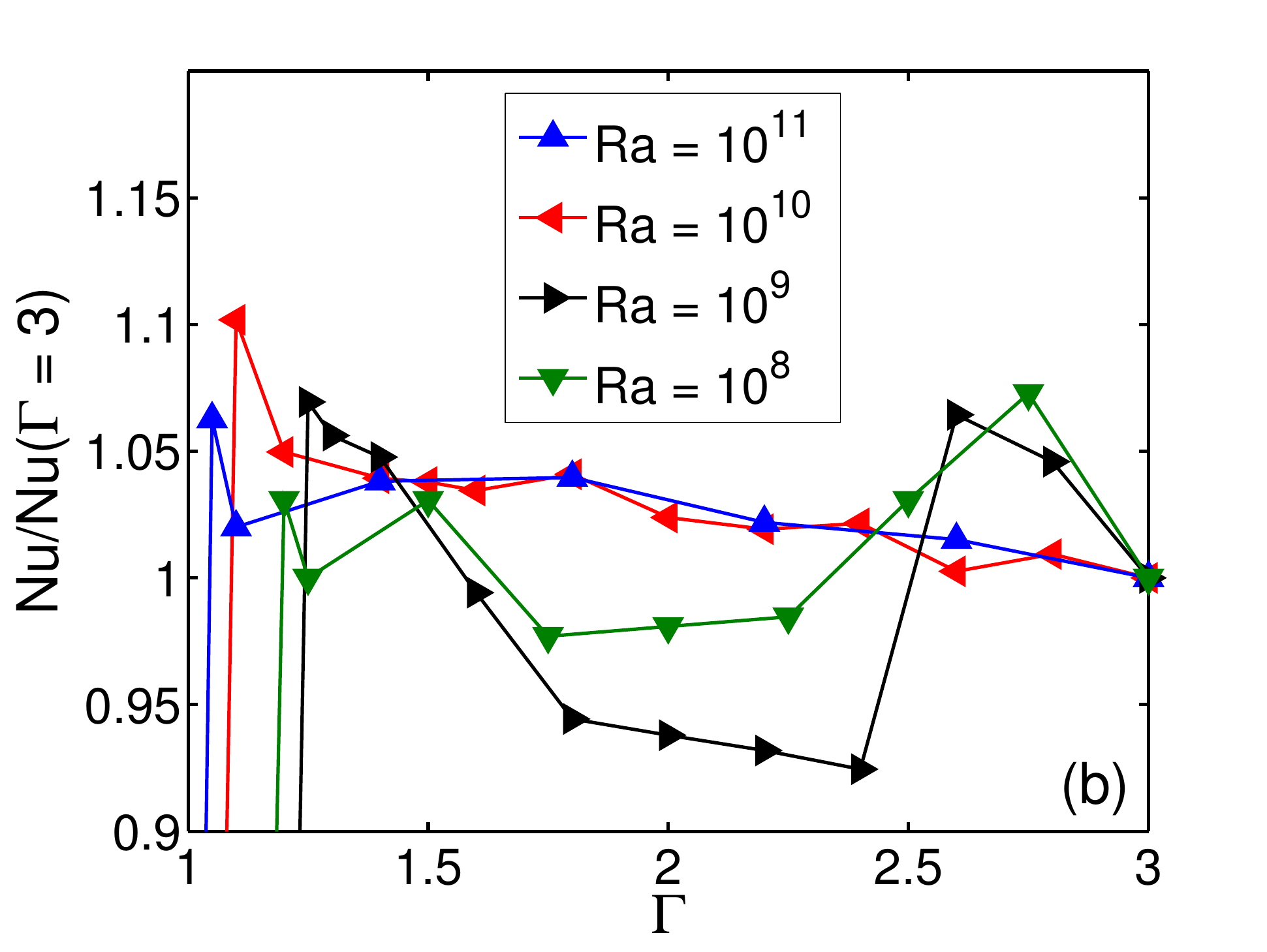}}
\caption{(Color online) a) Nu vs $\Gamma$ for periodic sidewalls and no-slip plates for $\tra \in \{10^8,10^9,10^{10},10^{11}\}$, see legend in b). Jumps in Nu can be seen around $\Gamma \approx 1.25$, $\Gamma \approx 1.25$, $\Gamma \approx 1.10$ and $\Gamma \approx 1.05$ for $\tra = 10^8$, $\tra = 10^9$, $\tra = 10^{10}$ and $\tra = 10^{11}$, respectively. The Nusselt number statistics at lower $\Gamma$ than the jump are more difficult to converge than the other data points due to the bursting nature of $\tnu(t)$ and have a larger error. They are included to indicate the change of flow state. In b) Nu is compensated with $\tnu(\Gamma=3)$.}
\label{fig:PD2}
\end{figure*}

The observation that the dependence of Nu on $\Gamma$ differs between no-slip and free-slip sidewall boundary conditions can be rationalized as follows. In the stress-free case, the aspect-ratio $\Gamma = D/H$ can be interpreted as the ratio between the proportion of the enclosing wall that draws energy out of the system (no-slip plates) and the proportion that has no effect on the flow (stress-free sidewall). With decreasing $\Gamma$, this ratio decreases and thus there is less drag experienced by the flow at lower $\Gamma$, resulting in a higher Nu for stress-free sidewalls. In case of no-slip sidewalls, following the same reasoning, one expects that the Nu does not depend on $\Gamma$ provided that the flow structure remains the same. The difference in Nu for no-slip sidewall between thin and square ($\Gamma = 1$) cells must then be a result of a changing flow state. While in $\Gamma = 1$ cells a single roll is present, in $\Gamma = 0.33$ cells there are four vertically stacked rolls, see figure \ref{fig:PD3}b. In 2D, the heat flux through the system depends on the amount of vertically stacked rolls \cite{poe11}, which explain the difference in Nu.

There are exceptions to this trend, as there are specific cases in which an aspect-ratio reduction resulted in an increased heat flux for no-slip walls \cite{hua13}. In addition, for the no-slip sidewall data seen in figure \ref{fig:FSNS}, Nu for $\Gamma = 0.33$ is lower than Nu for $\Gamma = 1$ for most of the Ra range explored. However, at $\tra = 4.64 \times 10^{10}$ this is opposite, which has been attributed to the break up of the large scale rolls in the flow \cite{poe12}.

We can further quantify the trend of $\tnu(\Gamma)$ for thin cells and stress-free sidewalls, and compare against the no-slip case. For no-slip sidewalls, $\tnu(\Gamma)$ shows relatively large jumps between vertically stacked roll states with Nu generally invariant with $\Gamma$ (for $\Gamma < 1$) provided that the roll state does not change \cite{poe11}. In figure \ref{fig:FS}a the corresponding results for stress-free sidewalls can be seen. In contrast to the no-slip case, no jumps can be seen and Nu decreases monotonically with $\Gamma$ up to $\Gamma = 0.95$. At $\Gamma \approx 1$ the flow transitions from a roll-state with one to a roll-state with two horizontally aligned rolls. There appears to be multistability in this transition region. The absence of jumps for $\Gamma < 1$ can be directly attributed to the fact that for $\Gamma \leq 0.95$ the system is exclusively in a single roll state while for $\Gamma = 1$ the system is in a roll state with two horizontally arranged rolls, resulting in a jump in Nu. 

\subsection{Wide cells}

\begin{figure*}
\centering
\subfigure{\includegraphics[width=0.45\textwidth]{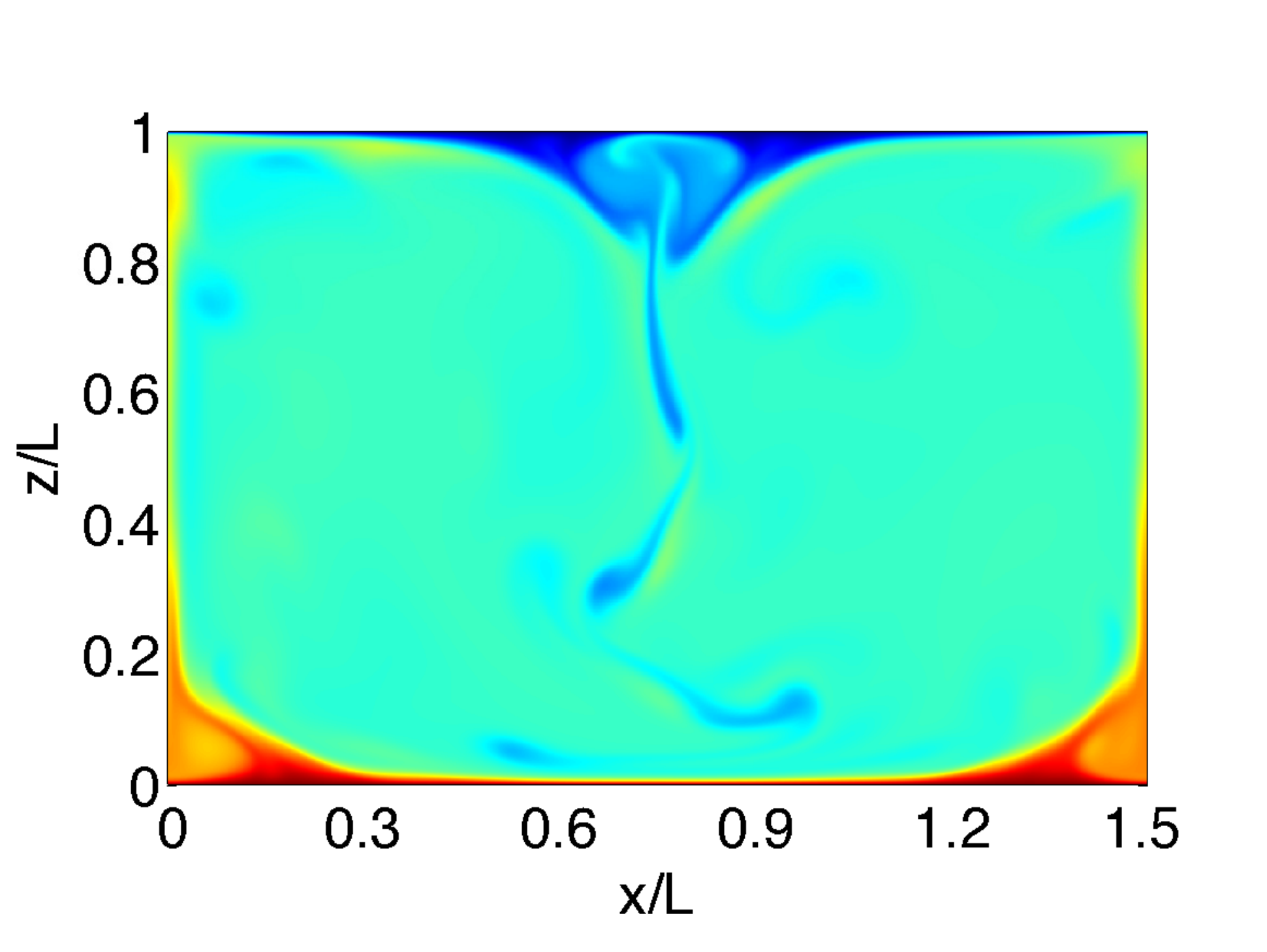}}
\subfigure{\includegraphics[width=0.45\textwidth]{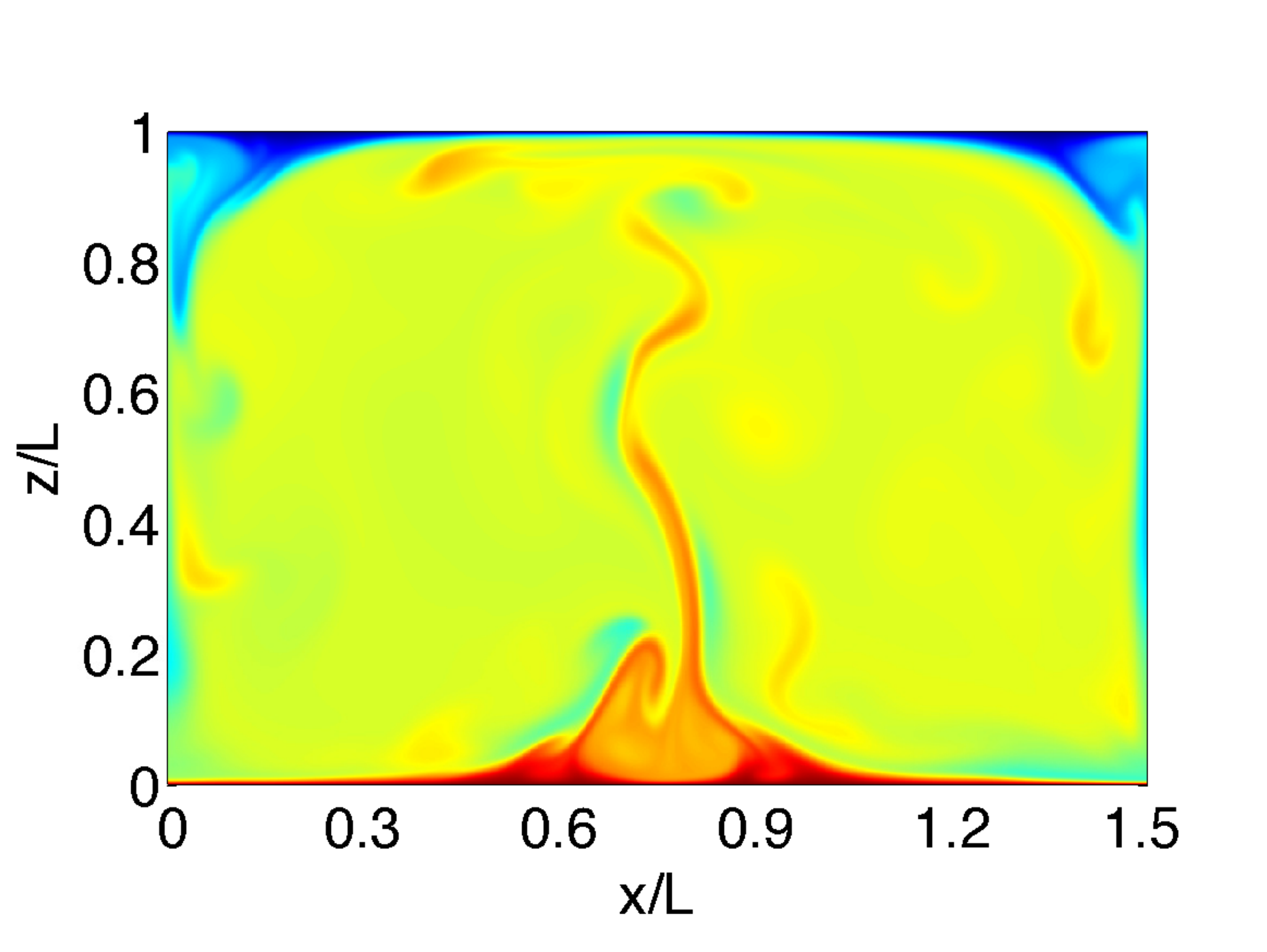}}
\caption{(Color online) Temperature field snapshots for $\tra = 2.15\times 10^8$ and $\Gamma = 1.5$ with no-slip plates and stress-free sidewall boundary conditions of two different statistically steady states. The colormap ranges from $0 \le \theta \le 1$. The area averaged mean temperature $\langle \theta \rangle_{A,t}$ is $0.44$ (left panel) and $0.56$ (right panel).}
\label{fig:SFCOLDHOT}
\end{figure*}

Small $\Gamma$ cells are dominated by the presence of the sidewall. However, even if one could expect this dependence to progressively vanish for the case of large $\Gamma$, Nu for 2D RB systems with no-slip sidewalls has a strong $\Gamma$ dependence up to very large values of $\Gamma$ \cite{poe12} (i.e. $\Gamma \lesssim 26$ for $\tra=10^8$). Here we repeat the analysis of ref. \cite{poe12}, but now for cells with lateral periodicity, to check whether the $\Gamma$ dependence is still present at such high values of $\Gamma$ and study the scaling of Nu(Ra).

In figure \ref{fig:FS}b Nu against Ra for laterally periodic and stress-free sidewalls are compared for $\Gamma = 1.5$ and $\Gamma = 2$. Nu for stress-free sidewalls is higher than for periodic sidewalls. This difference can be attributed to two effects, namely the roll state imposed by the impermeable walls and the drag from the sidewall. The amount of horizontally stacked rolls for periodic sidewalls, in contrast to impenetrable sidewalls (NS and SF sidewalls), is restricted to even numbers. At $\Gamma = 2$, for the studied parameter range, the periodic simulations showed two horizontally stacked rolls while the stress-free simulations showed three rolls. For both simulations, the initial conditions consisted of two rolls, to ensure that the heat flux is not affected by a different roll state. 

Nu heavily depends on the amount of horizontally stacked rolls \cite{poe11} and this effect most likely contributed to the difference in Nu between the evaluated cases. However, there could also be other causes of the difference in Nu. In order to find out, $\Gamma = 1.5$ was simulated for the same Ra as for $\Gamma = 2$. At this aspect-ratio both laterally periodic and stress-free sidewall simulations have a roll state consisting of two rolls. It can be seen that the difference in Nu between PD and SF is smaller, as expected, and that the difference in stress at the boundaries alone results in a lower Nu for laterally periodic cells as compared to stress-free sidewalls. 

The figure shows that Nu becomes $\Gamma$-independent for both BC's at sufficiently large Ra. This finding adds the conclusion that not only laterally periodic cells lose the $\Gamma$ dependence, but that this is also valid for stress-free sidewalls at $\Gamma\approx2$ for $\tra \ge 10^{10}$. We now focus on the aspect-ratio dependence of periodic sidewall simulations at different Ra.

Periodic boundary conditions are used to approximate infinite aspect-ratio, i.e. an absence of finite size effects, on a finite size domain. Thus, the aspect-ratio in periodic sidewall simulations does not reflect a physical domain but instead the period size of the computational domain. This aspect-ratio cannot be arbitrarily small due to two reasons: i) there must always be an even number of horizontal rolls in a periodic system, and ii) wavelengths larger than the domain are not simulated and thus not accounted for. However, as the aspect-ratio is proportional to the computational cost, it is of interest to study the minimum aspect-ratio $\Gamma_{\text{min}}$ that still produces physical solutions. When considering three dimensional (3D) flows with a large scale circulation (LSC), periodicity can be imposed in two distinct dimensions. The periodic dimensions can either be parallel or perpendicular to the vorticity vector of the LSC. It is expected that parallel periodicity has less effect on the flow than perpendicular periodicity. Parallel periodicity is often used in Taylor-Couette simulations \cite{ost13}, where, unlike in 3D RB \cite{bro05b}, the orientation of the LSC is known {\it a priori}. In 2D RB this periodicity is unavoidably imposed in the perpendicular direction and thus a strong effect is expected.

\begin{figure*}
\centering
\subfigure{\includegraphics[height=0.37\textwidth]{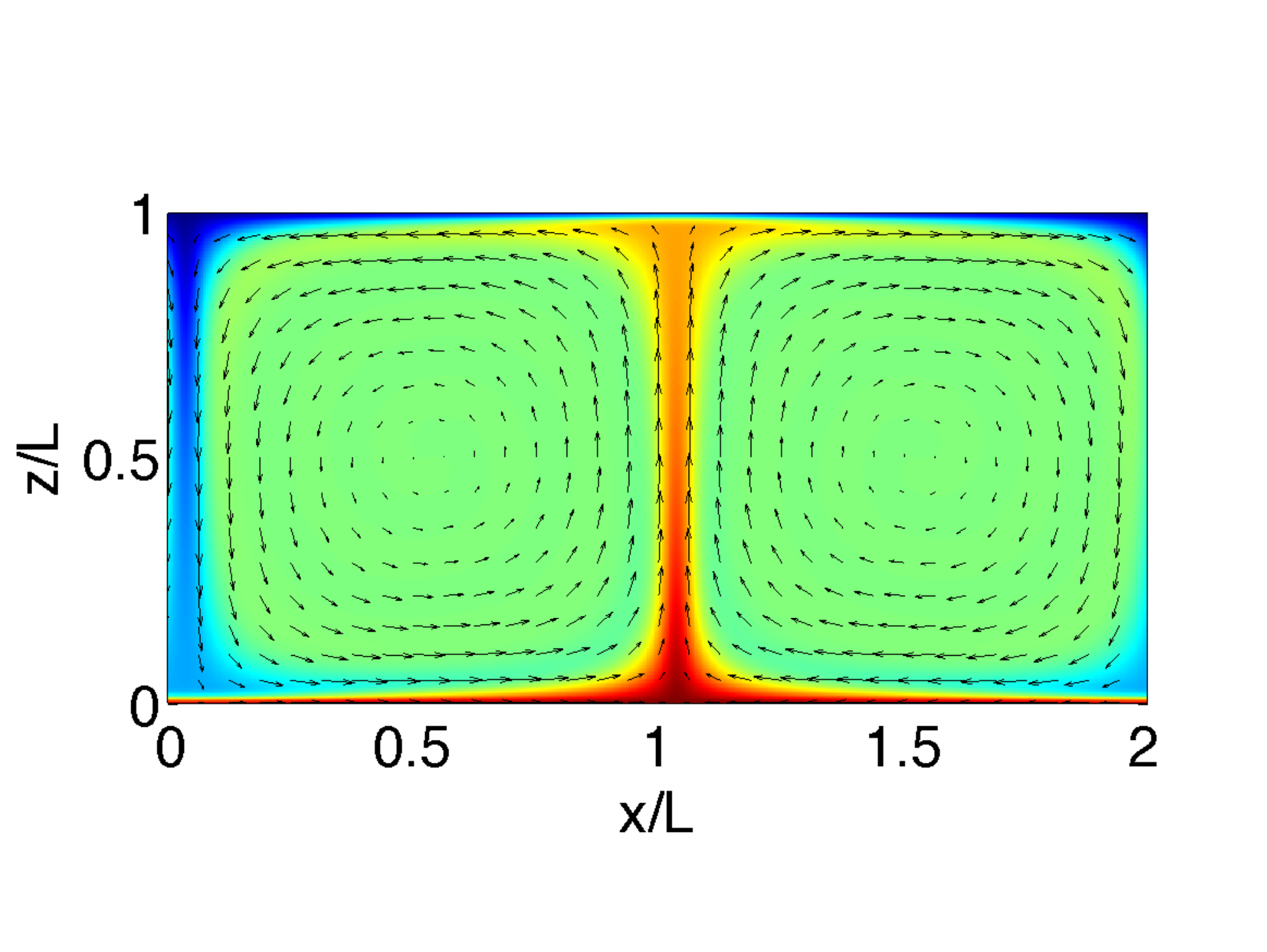}}
\subfigure{\includegraphics[height=0.37\textwidth]{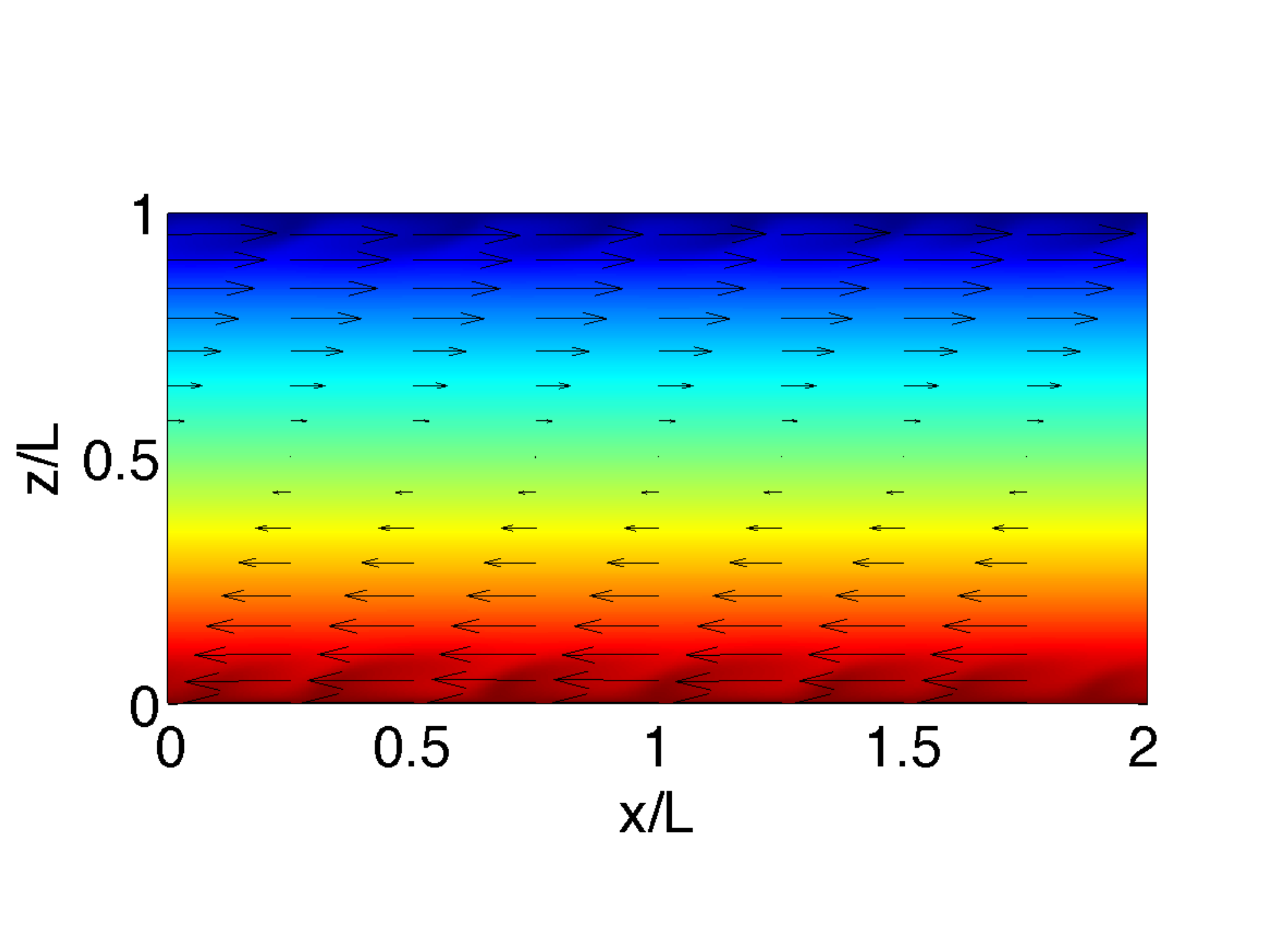}}
\caption{(Color online) Temperature field snapshots with velocity vectors superimposed for stress-free plates and periodic sidewalls for $\tpr = 1$ and at $\tra = 10^6$ (left panel) and $\tra = 10^8$ (right panel) in a $\Gamma = 2$ cell. The temperature ranges from $0 \le \theta \le 1$. The size of the arrows is proportional to the absolute velocity. The flow is roll-like in a) and zonal in b). Corresponding movies can be found in the supplemental material.}
\label{fig:zonalsnaps}
\end{figure*}

In figure \ref{fig:PD2} Nu as function of the period $\Gamma$ is shown for four different Ra. One would expect that $\Gamma_{\text{min}} \approx 2$ as the minimum number of rolls is two and the aspect-ratio of one roll is approximately unity \cite{poe12}. It is therefore surprising that a weak dependence of $\tnu$, i.e. $\leq 5\%$, on $\Gamma$ is seen from $\Gamma_{\text{min}}$ significantly lower than two, where $\Gamma_{\text{min}}$ is indicated by the jump in Nu(Ra). For $\tra=10^9$ the aspect-ratio dependence is larger than for higher Ra. This is in particular visible in figure \ref{fig:FS}b. At $\tra=10^9$, Nu differs approximately 8\% between $\Gamma=1.5$ and $\Gamma=2$, while this difference vanishes at $\tra=10^{10}$. $\Gamma_{\text{min}}$ appears to decrease for increasing Ra. We expect this to be a result of a weaker LSC \cite{poe13} and decreased structure size at higher Ra.

We wish to highlight a peculiarity observed for stress-free sidewall and no-slip plates. Namely, that the mean temperature can be $\langle \theta \rangle_{A,t} \not= 0.5$. This can be seen in more detail in figure \ref{fig:SFCOLDHOT}. Even though there is top bottom symmetry in the boundary conditions and in the equations there is a deviation from the expected value $\langle \theta \rangle_{A,t} = 0.5$. As the system is not turbulent enough, it might take infinite time for the system to explore all possible states and the symmetry can be sponteneously broken due to the different flow states. The system is in a horizontally stacked double roll state where the hot plumes are moving up at the center of the no-slip plate and the cold plumes move down along the stress-free sidewalls. In this case $\langle \theta \rangle_{A,t} > 0.5$, as the hot plumes are well mixed by turbulence while the cold plumes seem to lose their heat through molecular diffusion, see figure \ref{fig:SFCOLDHOT}a. A similar but opposite effect can be obtained in other realizations of the same system with different initial conditions. In steady state axisymmetric convection with stress-free plates and sidewalls \cite{ume88} the core temperature is smaller than $0.5$ in case of a rising central plume. \\
We have checked the dependence of the deviation of the mean temperature from $0.5$ on the central plume location by performing the transformation $u_i(x,z) \mapsto u_i(x,L-z)$ and $\theta(x,z) \mapsto 1 - \theta(x,L-z)$ on the hot solution of $\Gamma = 1.5$, $\tra = 2.15 \times 10^8$. This transformation flips the velocity and temperature fields upside-down and inverts the temperature. While before the transformation in statistically steady state $\langle \theta \rangle_{A,t} = 0.54 > 0.5$, converging the simulation with the transformed field results in $\langle \theta \rangle_{A,t} = 0.46 < 0.5$, of which a snapshot can be found in figure \ref{fig:SFCOLDHOT}b. This indicates that the deviation from $0.5$ is a flow state effect heavily dependent on the initial condition. Even though the time and volume averaged heat flux through the system is identical in both cases, the area-averaged and local temperature strongly differ between these two flow states.

\section{Zonal flow}

\subsection{Stress-free plates}
\label{sec:SFP}

Up to now we have focused on the case with no-slip plates and varied the sidewall boundary conditions. Of course, another degree of freedom of the boundary conditions are the horizontal plates. Using a stress-free boundary condition instead of no-slip results in huge differences in both flow topology and heat flux. 
Stress-free plates are used in theory to simplify boundary conditions, as well as infinite (periodic) systems. This BC is a critical requirement for theorems that use the vanishing vorticity condition at the plates, which goes all the way back to original analysis by Rayleigh, but it is also used for more recent work, such as the finite-Pr upper bound theorem of \cite{whi11} that gives $\tnu \leq 0.2891\tra^{5/12}$. 

We performed simulations with (horizontal) stress-free plates and used lateral periodicity with $\Gamma=2$ to approximate an infinite system. For $\tra \gtrsim 10^7$ the flow topology consists of two shearing layers with predominantly horizontal motion. The flow at the lower half of the domain moves in opposite direction as the top. This is because the total momentum must be conserved due to the absence of momentum transport to the boundaries. The flow is in a state for which Nu shows bursts with long ``quiet'' intervals where $\tnu(t) \approx 1$. This flow state is referred to as {\it zonal flow} \cite{tho70,kri81,ruc96,gol14}, similar to the zero-wavenumber flows observed in toroidal plasmas and planetary atmospheres, e.g.\ on Jupiter. The strong horizontal motion is driven by the so-called ``shearing mechanism''. Plumes emitted from the one plate move vertically at first until they are deviated horizontally by the opposite plate, driving the horizontal motion. 

\begin{figure*}
\centering
\subfigure{\includegraphics[height=0.28\textwidth]{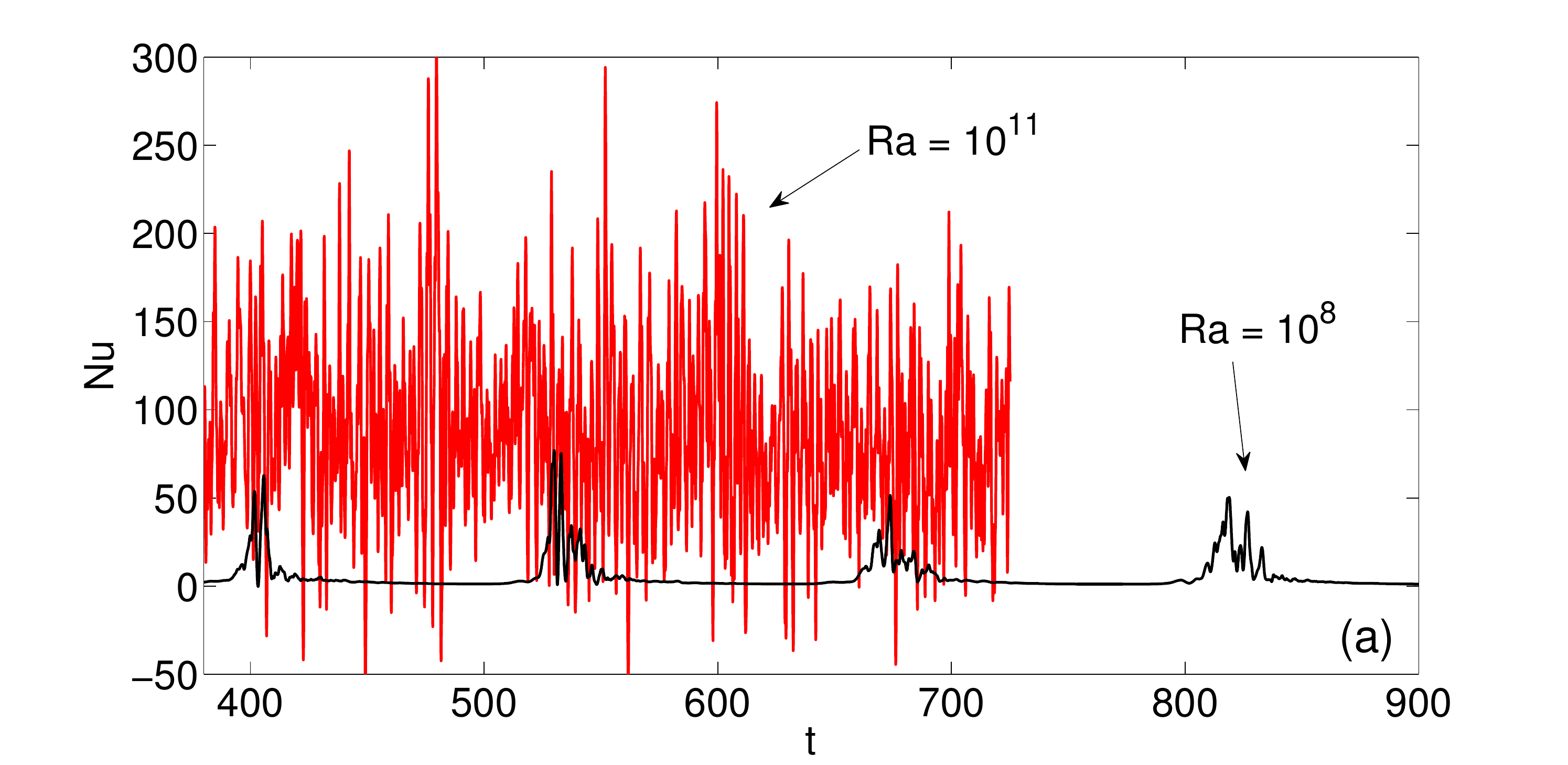}}
\subfigure{\includegraphics[height=0.28\textwidth]{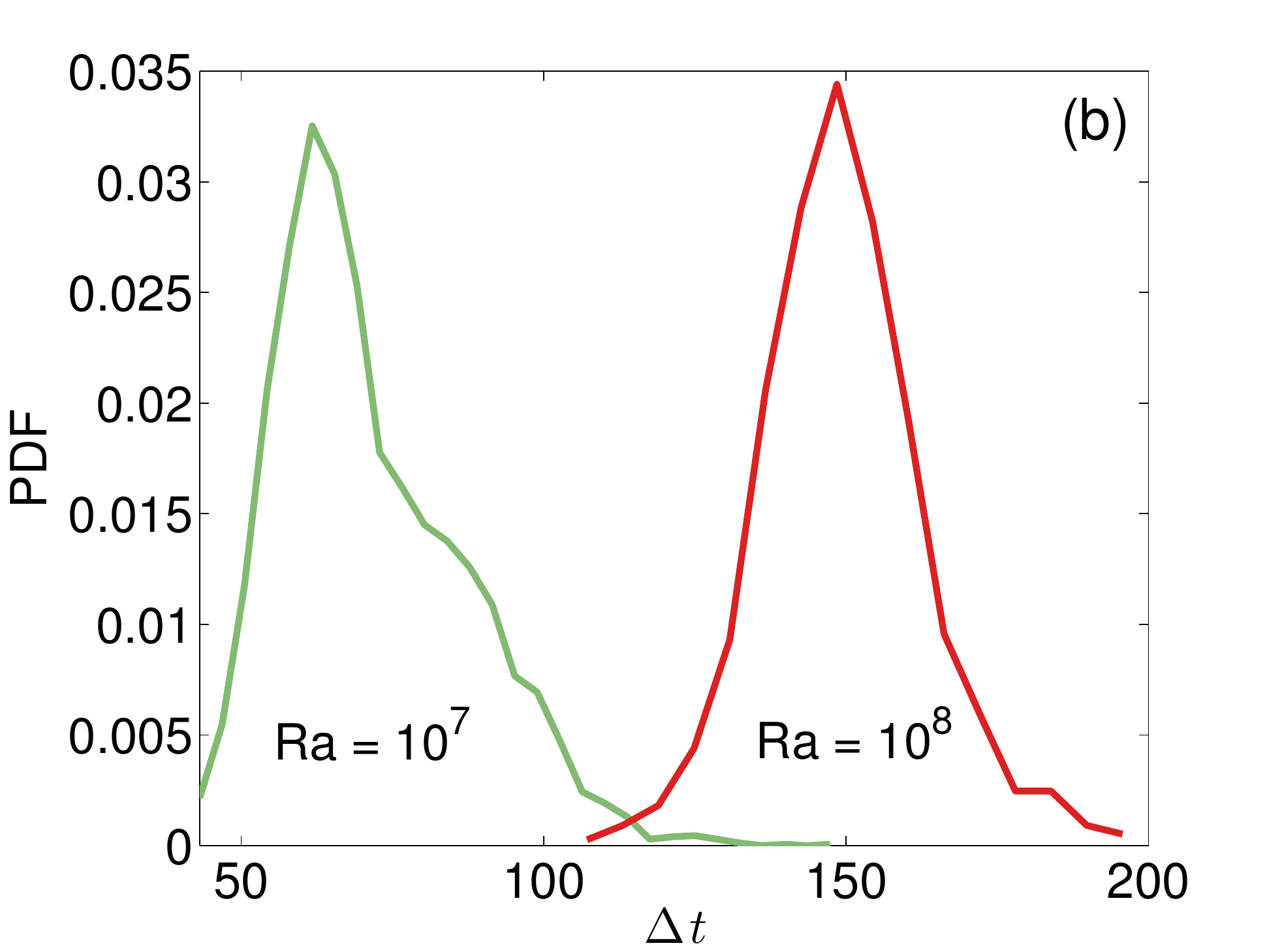}}
\caption{(Color online) a) Nu as a function of dimensionless time in freefall time units at $\Gamma = 1/2$, for $\tra = 1\times10^8$ and $\tra = 1\times10^{11}$, NS plates and PD sidewalls. At $\tra = 10^8$ we observe bursting behaviour while at $\tra = 10^{11}$ we do not: The zonal flow is bursting for lower Ra and sustained for higher Ra. Note the large difference between the time scale of the bursts at $\tra = 10^{8}$ and the fluctuations at $\tra = 10^{11}$. b) Probability density function of the time interval between the bursts $\Delta t$ for $\tra = 10^7$ and $\tra = 10^8$.}
\label{fig:PD4}
\end{figure*}

Depending on Pr, there is a threshold Ra beyond which the flow transitions from roll-like to zonal flow. Hysteresis is observed at this transition \cite{gol14}. In figure \ref{fig:zonalsnaps} snapshots can be seen of the flow below and above this threshold at $\tpr = 1$. At $\tra = 10^6$ the flow is still roll-like; a large scale circulation with accompanying plumes. Increasing Ra to $\tra = 10^8$ the flow transitions to a zonal flow. In the corresponding snapshot in figure \ref{fig:zonalsnaps}b the flow is in the $\tnu(t) \approx 1$ interval between bursts, in which a linear temperature gradient can be seen with strong horizontal flow. This interval phase is interrupted by large bursts in Nu(t). During these burst phases there is high mixing before the interval phase is re-initiated. Zonal flow is only observed in two dimensions. In our three-dimensional simulations using SF plates, periodic sidewall and a cartesian geometry with $\Gamma = 1$, no zonal-flow has been observed for $\tra \in \{10^7,10^8,10^9\}$. Bursts in Nu have been observed for wind-reversals \cite{sug10} although in that case the bursts are not preceded by ``quiet'' i.e. $\tnu \approx 1$ flow states. 

The zonal flow state is detrimental for obtaining Nu(Ra) scaling as Nu is close to one and, due to the intermittent nature of the bursts, time-averaged quantities are very hard to converge, requiring simulations of several tens of thousands of large eddy turnover times. In order to obtain a converged Nu to see how the system approaches the theoretical upper bounds, we prevent zonal flow by increasing Pr as this increases the shear between the bottom and top layer. Increasing to $\tpr = 40$ prevents zonal flow for $\tra < 10^{10}$. In the low Ra regime the flow is composed of two rolls with very thin plumes streaks in between, very similar to infinite Pr flows with SF plates \cite{pet11}. The absolute Nu for a given Ra is nearly double of that in the case of no-slip plates. However, the Nu(Ra) scaling exponent is similar, namely 0.28,  much smaller than the upper bound exponent of $5/12 \approx 0.42$ \cite{whi11}, rigorously derived for these boundary conditions. 

\subsection{Suppressed periodicity}

\begin{figure*}
\centering
\subfigure{\includegraphics[width=0.32\textwidth]{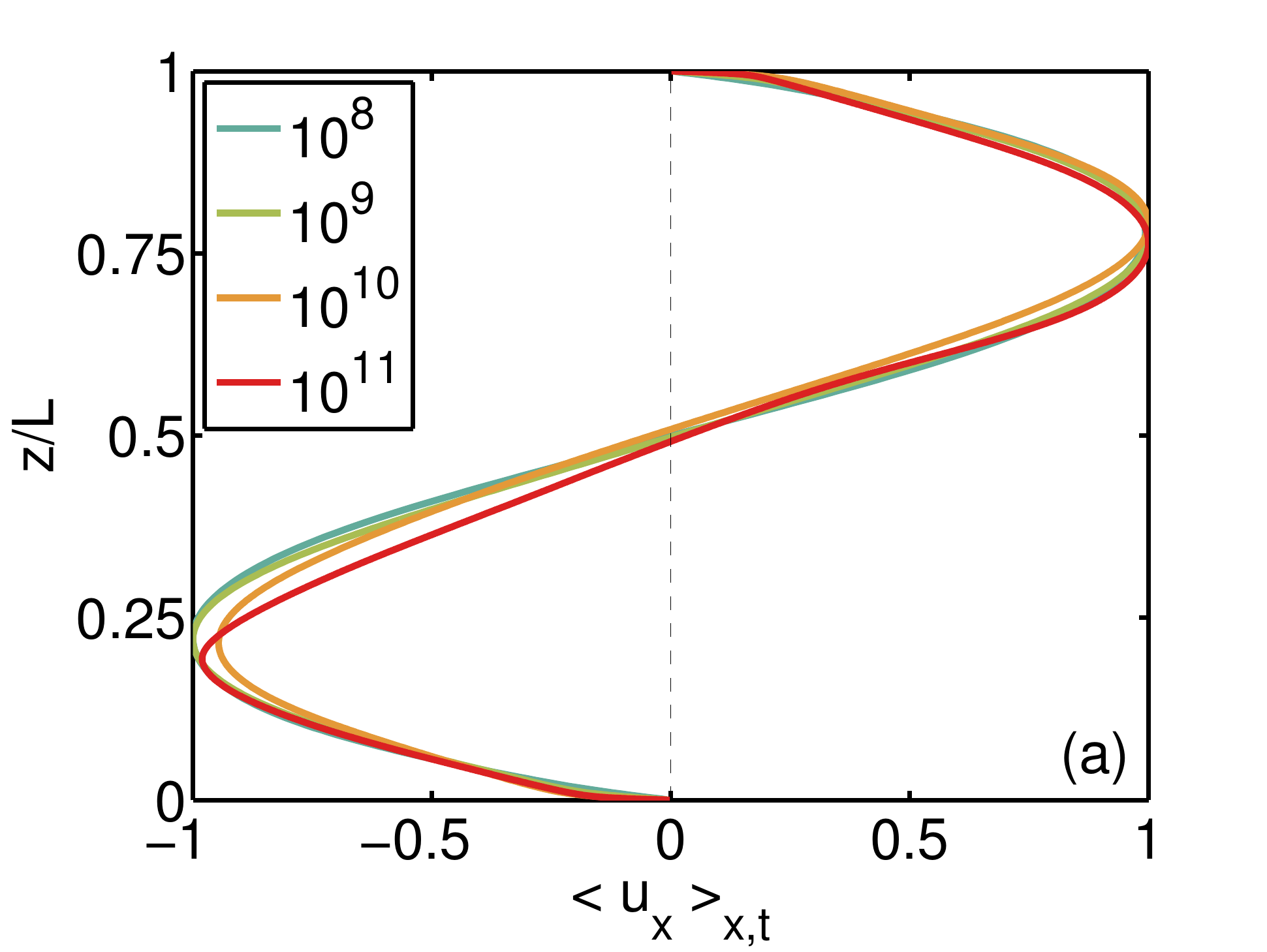}}
\subfigure{\includegraphics[width=0.32\textwidth]{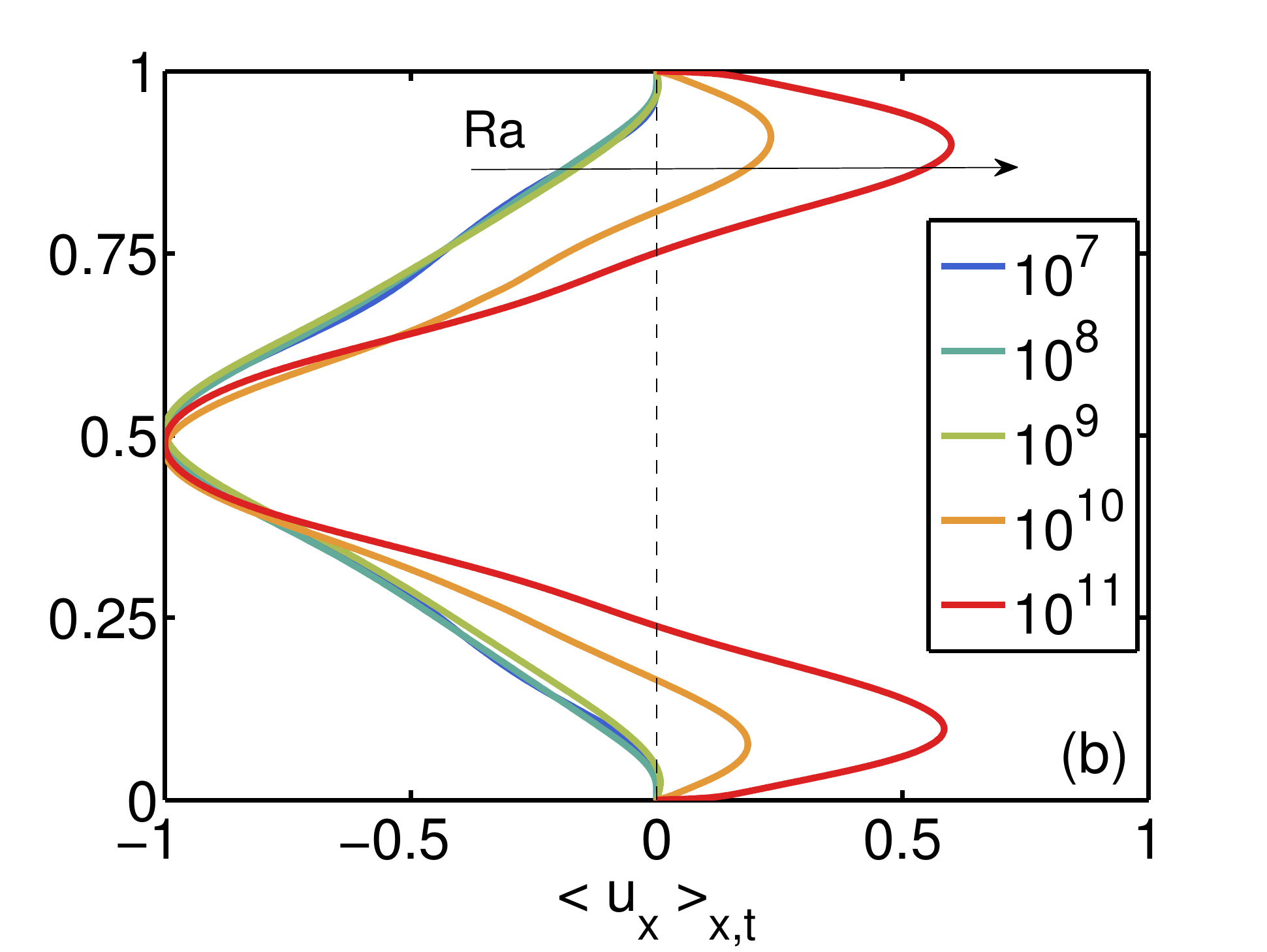}}
\subfigure{\includegraphics[width=0.32\textwidth]{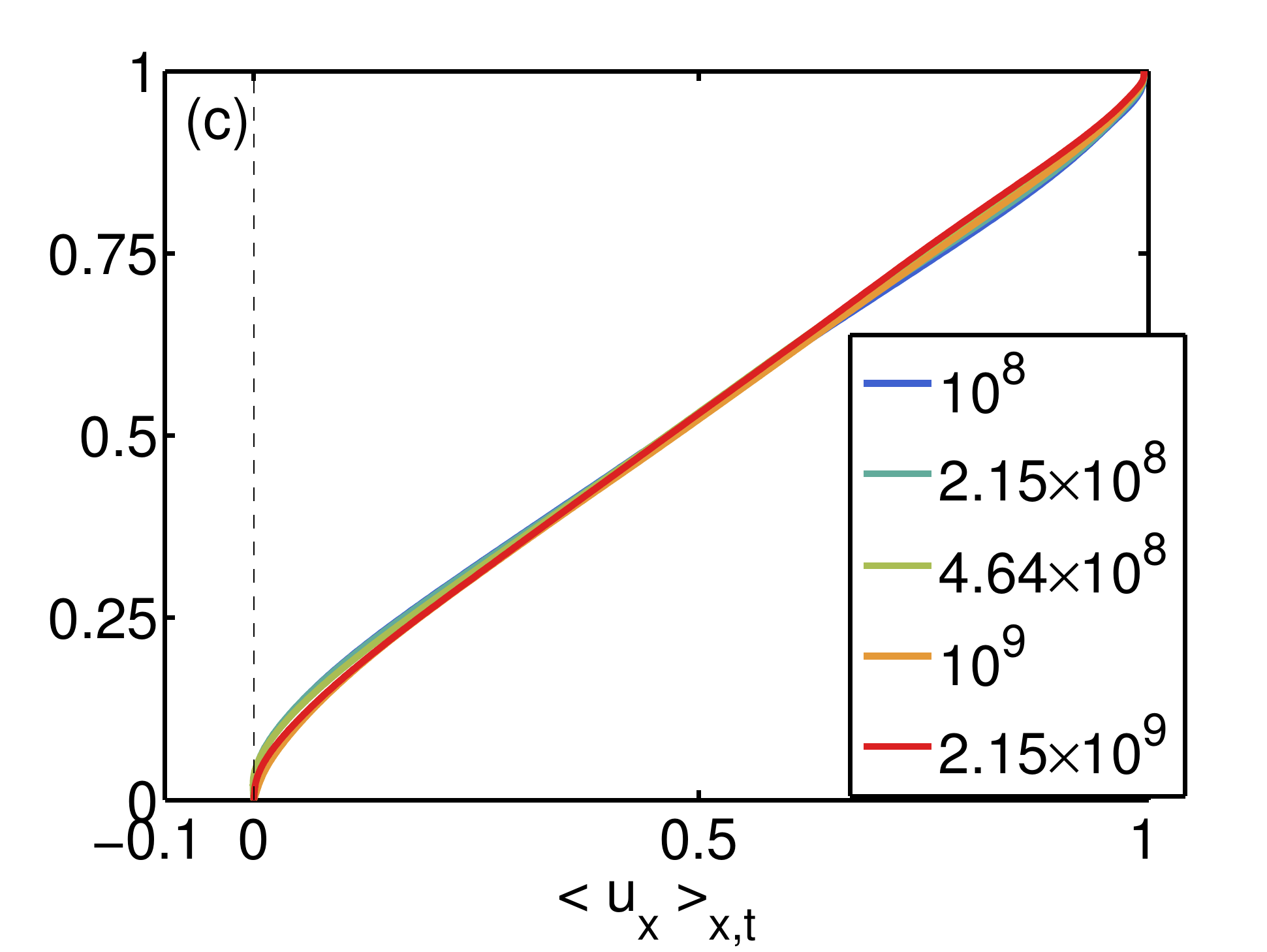}}
\caption{(Color online) Mean horizontal velocity $\langle u_x \rangle_{x,t}$, normalized by its maximum, as a function of vertical position $z/L$ for different Ra (see legend) in a) $\Gamma = 1$ cells with lateral periodicity and no-slip plates, b) $\Gamma = 0.5$ cells with lateral periodicity and no-slip plates and c) $\Gamma = 2$ cells with lateral periodicity and asymmetric plates.}
\label{fig:PD5}
\end{figure*}

Zonal flow is not inherent to the stress-free plate boundary condition. It is also observed for no-slip plates when $\Gamma$ is too low for the lateral periodicity and the flow becomes dominated by artifacts. For $\Gamma < \Gamma_{\text{min}}$ the flow is zonal for all the studied Ra. First, we focus on $\Gamma = 1$, which is close to $\Gamma_{min}$ where the flow transitions from zonal to roll-like, see figure \ref{fig:PD2}. A strong characteristic of zonal-flow as compared to roll-like flow is that the mean horizontal velocity $\langle u_x \rangle_{x,t}$ as a function of $z/L$ is non-zero \cite{gol14}. This can be seen in figure \ref{fig:PD5}a for $\Gamma = 1$. The two shearing layers are reflected in $\langle u_x \rangle_{x,t}$ by the two peaks with opposite velocity. The velocity profile seems nearly independent of Ra. Unlike with stress-free plates \cite{gol14}, the velocity in this case vanishes at the plates, inherent to the no-slip boundary condition. It thus appears that zonal flow is not intrinsic to stress-free plates. When the lateral period is too small to allow roll-like flow, the sideways deflected plumes have no freedom to return to their original plates. The plumes therefore have to move sideways, driving a zonal flow.

Decreasing the aspect-ratio to $\Gamma = 0.5$ increasingly suppresses the flow. In figure \ref{fig:PD4}a, Nu(t) for two different Ra is depicted. Nu(t) for $\tra = 10^8$ shows the Nu bursts with a peak value of approximately $Nu(t_{peak}) \approx 50$. For $\tra = 10^{11}$ the features of Nu(t) are completely different. Here, large symmetric fluctuations around the mean can be seen, characteristic for common RB systems. A similar transition is seen for zonal flow between stress-free plates \cite{gol14}. However, there the zonal flow type transitions from {\it bursting} to {\it sustained} when Pr is increased. Here, the transistion between the bursting and the sustained Nu is at $\tra = 10^{10}$ for constant $\tpr = 1$ and $\Gamma = 0.5$, with bursts found at lower Ra. At this same $\tra = 10^{10}$ the vertical profile of $\langle u_x \rangle_{x,t}$ starts to differ from the profiles at lower Ra (figure \ref{fig:PD5}b). For lower Ra, the horizontal motion is only in one direction, with a maximum at mid-height. At higher Ra, the velocity near the plates opposes the motion at mid-height and three regions of different horizontal motion are present. This shows that also for zonal flow there are different states, strongly dependent on the boundary conditions and correspondingly $\Gamma$.

The scale of the buoyant thermal plumes decreases for both increasing Ra and increasing Pr, allowing for more freedom compared to the fixed periodic length scale. In the $\Gamma = 0.5$ case, increasing Ra prevents the bursting flow state. In the sustained zonal flow state, there are no bursts in Nu. Most likely it is the heat buildup during the stagnant phase that results in the bursts. This is supported by the fact that for larger $\Gamma = 1$, the zonal flow is additionally sustained. The flowfield snapshot of $\tra = 10^{11}$ is displayed in figure \ref{fig:PD3}a. The flow topology in this snapshot is very different from typical RB (see e.g.\ figure \ref{fig:PD3}b), even though Nu(t) has a similar structure, its mean value is substantially lower, indicating that the suppression of heat flux still is significant at higher Ra. 

The interval between the bursts strongly depends on Ra. In figure \ref{fig:PD4}b, the probability density functions for $\tra = 10^7$ and $\tra = 10^8$ can be seen. The interval duration $\Delta t$ is extracted using a peak finding algorithm that finds local maxima. The minimum peak height and minimum peak distance are tuned by hand for each Ra. For increasing Ra, the mean interval increases while the variation appears approximately constant. For $\tra > 10^8$ the mean $\Delta t$ becomes so large that obtaining statistical convergence requires huge amounts of CPU time. It therefore remains unknown how the mean time interval between bursts increases up to the point that the bursting behaviour disappears. In addition no converged Nu can be obtained as a result of the increasing time interval.

These low $\Gamma$ PD sidewall simulations are the only cases we encountered with zonal flow when both plates are no-slip in two-dimensions. However, it has been observed in three-dimensional RB experiments in an annular cylinder \cite{kri81}, where all the enclosing boundary conditions are no-slip with periodicity introduced in the azimuthal direction. 

\subsection{Asymmetric plate boundary conditions}
\label{sec:AS}
The final variation we consider on the plates is using asymmetric velocity boundary conditions. These conditions are rarely studied, but have many applications; the ocean or a cooking pan can be thought of as having stress-free top at their free surfaces and a no-slip bottom. Asymmetric boundary conditions break the top-down symmetry in RB even in the Boussinesq approximation. In conjunction to asymmetric boundary conditions at the plates, we take a laterally periodic system, to avoid as much as possible sidewall effects. 

\begin{figure*}
\centering
\subfigure{\includegraphics[height=0.28\textwidth]{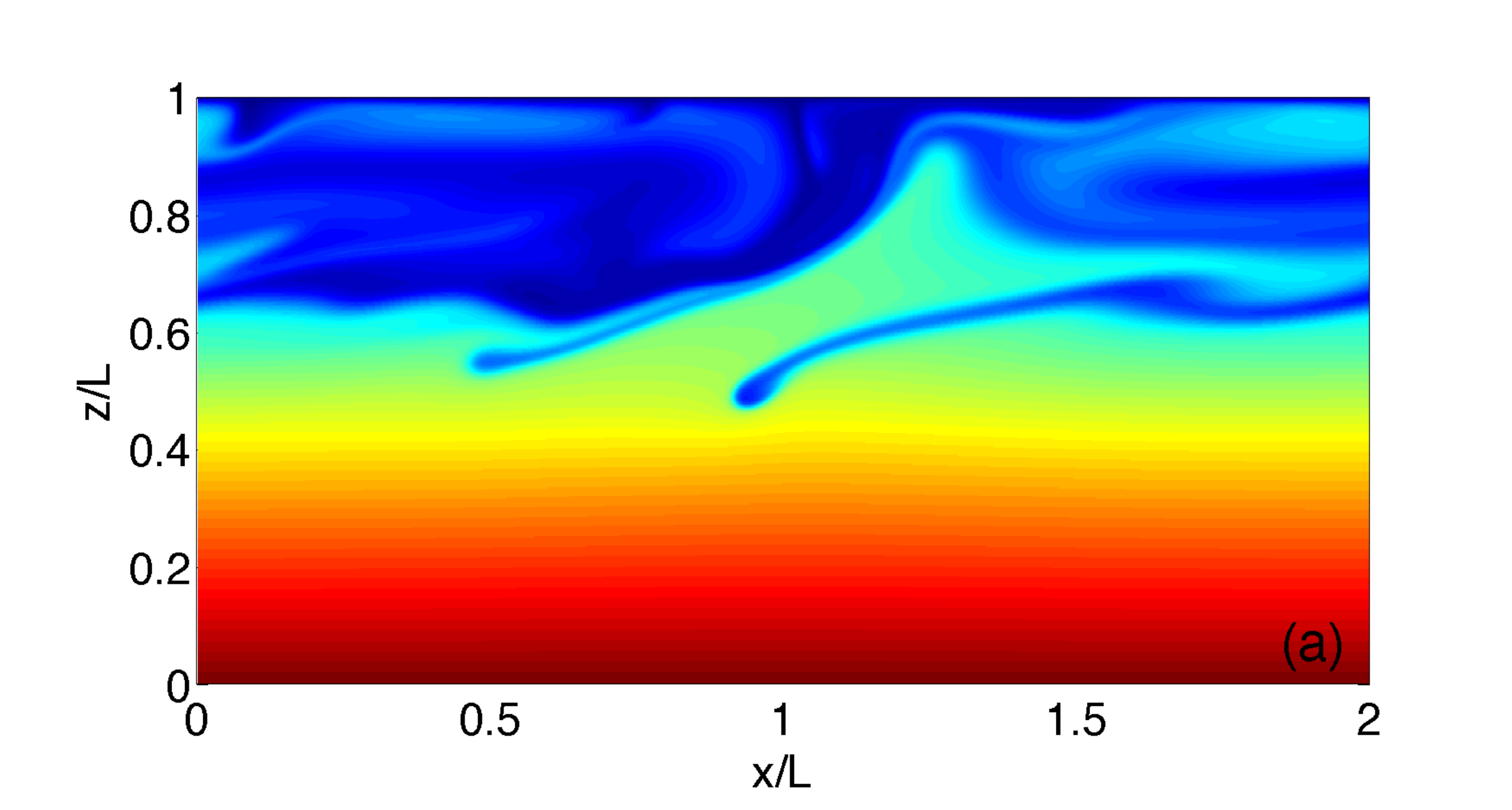}}
\subfigure{\includegraphics[height=0.28\textwidth]{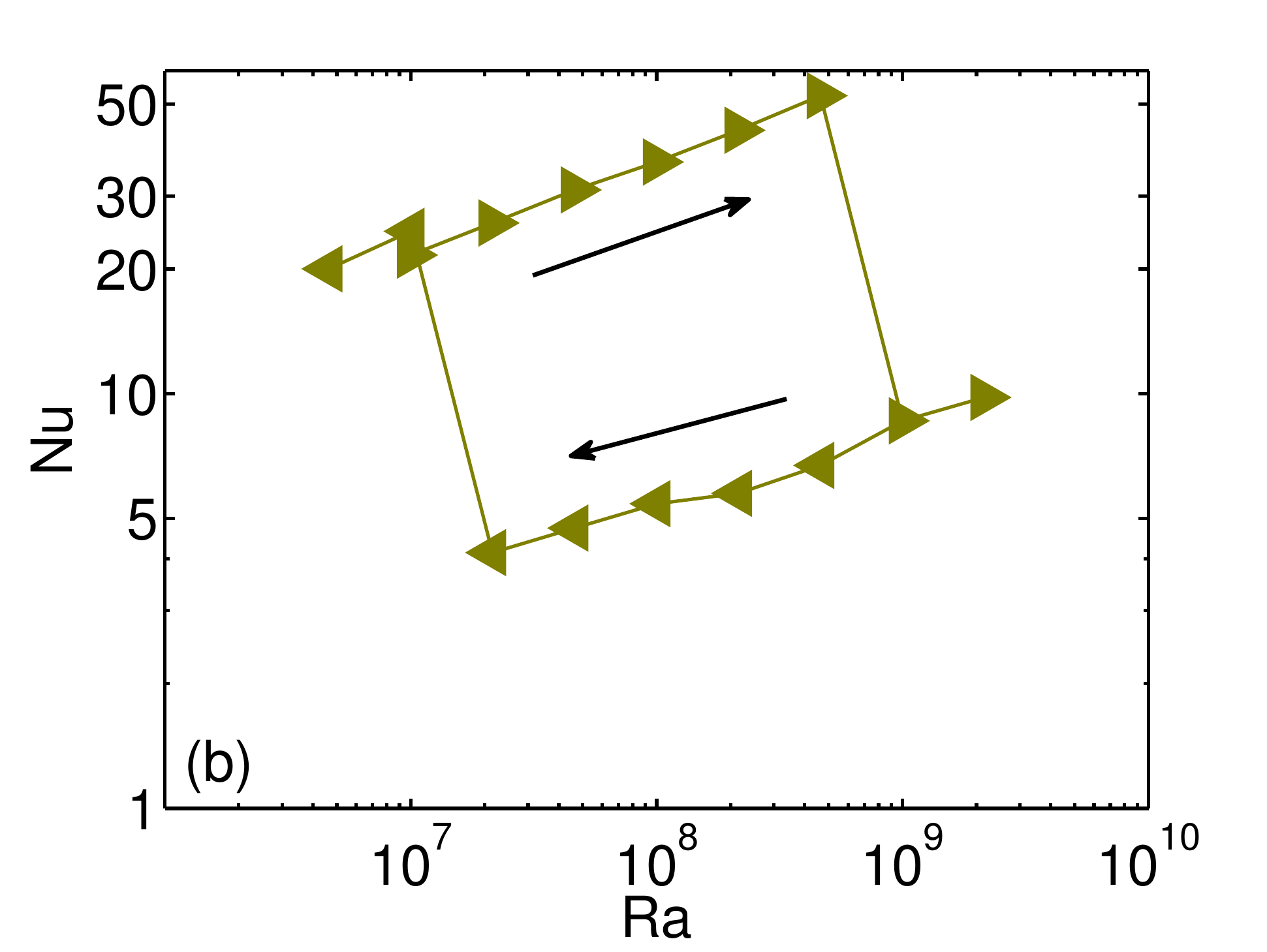}}
\caption{(Color online) a) Temperature field snapshot for asymmetric plates BCs and periodic sidewalls at $\tra = 4.64 \times 10^8$. The top boundary is stress-free and the bottom boundary is no-slip. . Movies of roll-like and zonal flow with asymmetric BCs can be found in the supplemental material. b) Nu as a function of Ra for asymmetric plate BC's and periodic sidewalls. Simulations where the initial conditions is taken from the lower Ra neighbouring data point are indicated with rightward-pointing triangles and where the initial condition is taken from higher Ra are indicated with leftward-pointing triangles.}
\label{fig:AS1}
\end{figure*}

Snapshots of the flowfield are displayed in figure \ref{fig:AS1}a. Here, it can be seen that we obtain partly zonal flow near the top plate where the BC is stress-free. Indeed, the system displays similar behaviour to the case where both plates are stress-free, namely bursts in Nu with $\tnu \approx 1$ intervals. The main difference between stress-free BCs and the asymmetric setup is obviously near the no-slip bottom plate. In symmetric zonal flow, the mechanism that drives zonal flow is conceptually understood as plumes from both plates driving the zonal flow at the opposite plate. However, the zonal flow present in this asymmetric setup has no plumes originating from the bottom plate, see \ref{fig:AS1}a. This shows that even without these plumes, the zonal flow in the upper, stress-free, layer can be driven by the interaction between the stationary layer at the bottom and the descending cold plumes alone.

In our simulations, we see this mixed zonal-flow state for $\tpr = 1$ and  $\tra > 10^9$. For lower Ra, the system displays large scale circulations without zonal flow, resulting in a more typical behaviour of Nu(t) fluctuating around the mean without periods where $\tnu \approx 1$. The thermal plumes in this Ra regime have a different shape; the plumes originating from the no-slip plate are surrounded with high vorticity while the others are not. The plumes that originated at the stress-free plate appear like long streaks that are less well mixed and have sharper gradients compared to their NS counterparts. 

Figure \ref{fig:AS1}b shows Nu as a function of Ra. As these Nusselt number data are very hard to converge the typical averaging time is at least one order of magnitude larger than for roll-like flows. The difference between Nu with and without zonal flow can be clearly seen. Moving upwards in Ra, i.e. using the solution field from a lower Ra simulation as an initial condition, the flow transitions at $\tra = 10^9$. Moving backwards, the transition back towards a flow without zonal flow is at $\tra = 10^7$. This difference indicates that there is hysteretic behaviour at this transition, similar to the double stress-free plates case \cite{gol14}. The vertical profile of $\langle u_x \rangle_{x,t}$ can be seen in figure \ref{fig:PD5}c for the cases with asymmetric plates that have zonal flow. Below the forward transition at $\tra = 10^9$, the horizontal motion is exclusively in one direction. For higher Ra, similar to the case described in the previous chapter, a layer is observed near the no-slip plate that has an opposite horizontal motion. At higher Ra, the plumes coming from the stress-free cold plate reach the no-slip plate and drive the zonal flow in the bottom of the cell, while at lower Ra the plumes do not reach that far into the flow and are dissipated before they can drive the zonal flow.

\begin{figure*}
\centering
\subfigure{\includegraphics[width=0.45\textwidth]{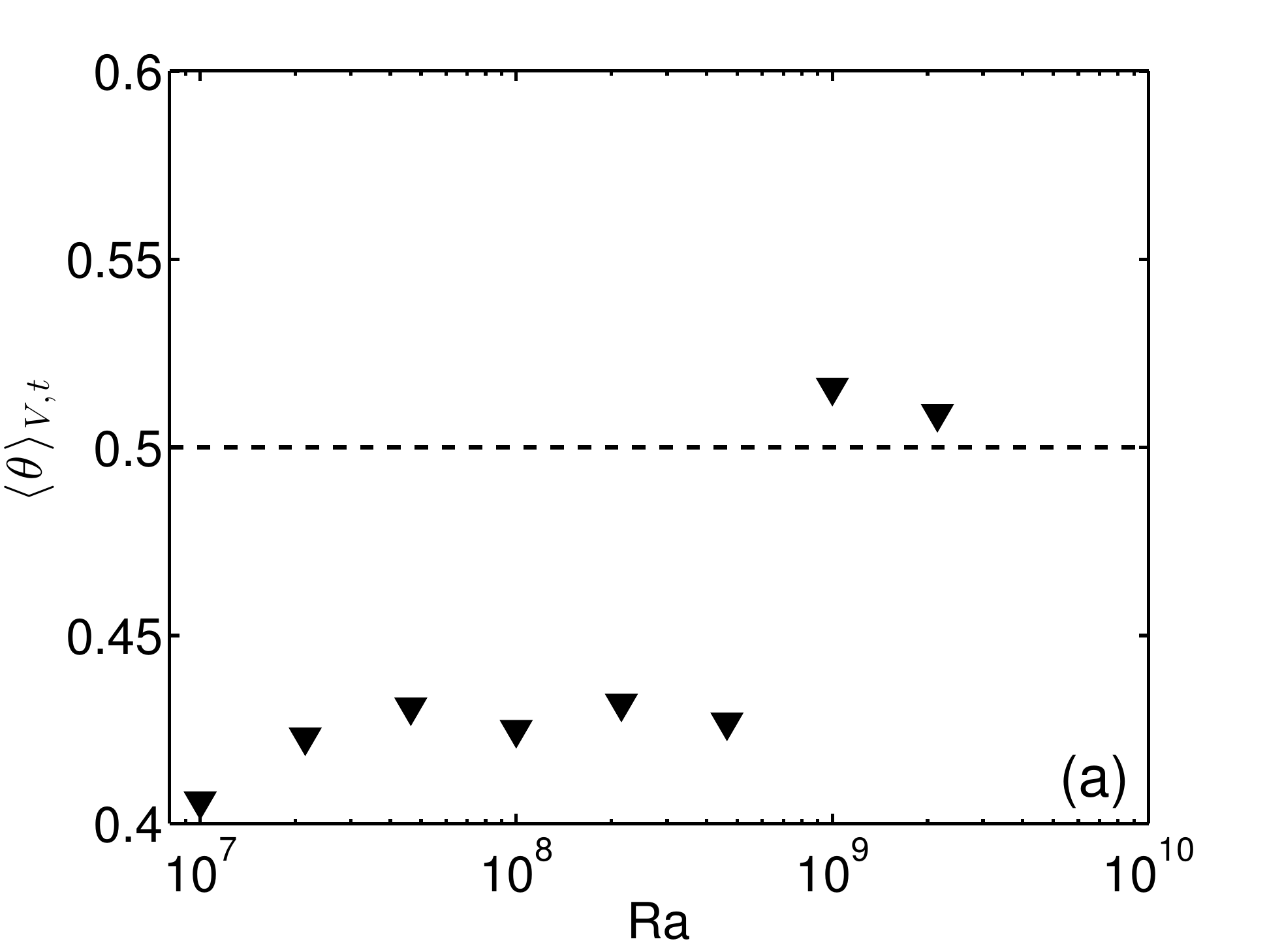}}
\subfigure{\includegraphics[width=0.45\textwidth]{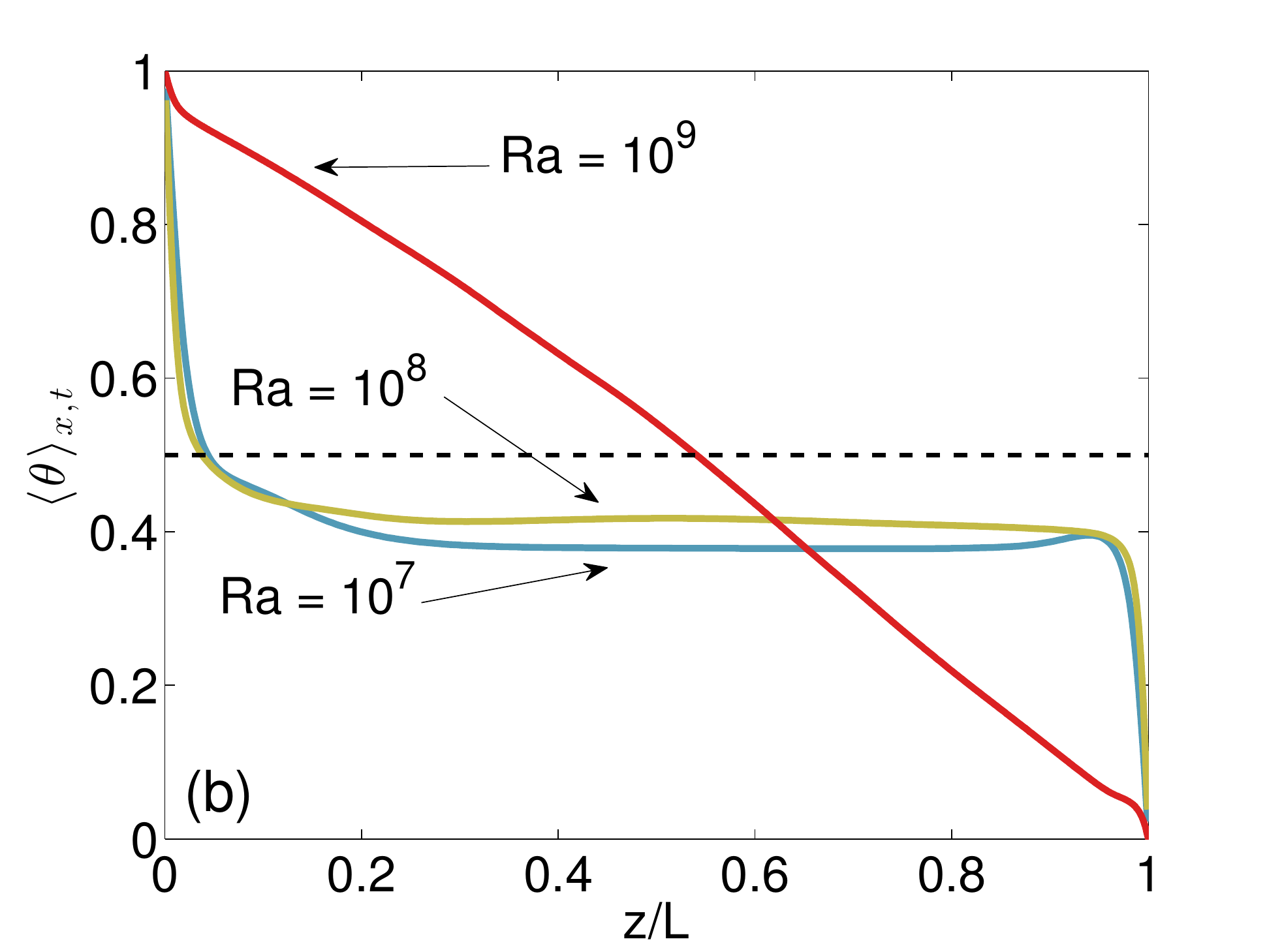}}
\caption{(Color online) a) The area -and time averaged temperature $\langle \theta \rangle_{A,t}$ as a function of Ra for asymmetric plate boundary conditions. Only the runs where the initial condition is taken from a lower Ra are included. b) The horizontal -and time average temperature $\langle \theta \rangle_{x,t}$ as a function of the vertical coordinate $z/L$ for $\tra = 10^7$, $\tra = 10^8$ and $\tra = 10^9$. The $\tra = 10^9$ data in this plot is taken from a simulation with zonal flow while the others are in a roll-like state. The dashed black line in indicates $\langle \theta \rangle_{x,t}=0.5$. For both panels, the asymmetry is introduced by making bottom plate (i.e. $z/L=0$) no-slip, while the top plate (i.e. $z/L=1$) is stress-free.}
\label{fig:AS2}
\end{figure*}

As the up-down symmetry is broken it is expected that the area and time averaged temperature $\langle \theta \rangle_{A,t}$ can deviate from the value of 0.5 commonly found in symmetric RB \cite{sug07}, similar to what is observed in non-Oberbeck-Boussinesq RB convection \cite{ahl06}. This quantity is plotted as a function of Ra in figure \ref{fig:AS2}a, where it can be seen that also within the Oberbeck-Boussinesq approximation the deviation from $0.5$ can be substantial, as seen here with asymetric plate boundary conditions. In addition, there is no clear dependence on Ra except for a large jump that corresponds to the transition between roll-like and zonal flow. It appears that for roll-like flow at lower Ra, the plate with the stress-free BC (in this case the cold plate) has a dominant effect on the bulk temperature, as $\langle \theta \rangle_{A,t} < 0.5$. The absence of the viscous boundary layer on this plate must therefore have a stronger effect on $\langle \theta \rangle_{A,t}$ than the increased vorticity at the no-slip plate. At the stress-free plate, the full thermal boundary layer is exposed to the LSC while at the no-slip plate it is (partly) nested in the viscous boundary layer. In contrast, in case of zonal-flow at higher Ra, there is a negligible deviation of the mean temperature from the symmetrical value of $0.5$.

In figure \ref{fig:AS2}b the vertical temperature profiles $\langle \theta \rangle_{x,t}$ as function of $z/L$ can be seen for various Ra. The no-slip and stress-free boundary layers are located near $z/L=0$ and $z/L=1$, respectively. For $\tra=10^7$ and $\tra=10^8$ the bulk profile is clearly colder than $0.5$. The two different boundary layer profiles are significantly different from each other for both these Ra. The no-slip plate boundary layer is less steep and larger than the stress-free one. A small overshoot in $\theta$ can be seen for $\tra=10^7$, which is typical for low Ra RB and not too small Pr. The boundary layer profile for $\tra=10^9$ deviates strongly from typical RB BL profiles \cite{zho10b}. It must be noted that this is a time average and so this profile is affected by both flow phases; the zonal flow phase and the burst. As there is negligible vertical motion in the zonal flow region, a near linear temperature profile can be seen, confirming that the convective heat flux is negligible and that the heat flux is dominated by molecular diffusion.

\section{Summary and Outlook}

As was seen in the previous sections, the velocity boundary conditions in two-dimensional Rayleigh-B\'enard convection have a large effect on the flow topology and consequently, the Nusselt number. This is in contrast with the temperature boundary conditions, which have been shown to have neglible effects when switching between isothermal and constant flux boundary conditions on the plates \cite{joh09}. 

The flow can be classified as roll-like and zonal. The roll-like flow state is composed of a large scale circulation interacting with thermal plumes. The resulting Nu(t) fluctuates around its mean. This flow state is most commonly observed when the horizontal plates are subject to a no-slip boundary condition, which is the case in experiments. The zonal flow state is however very different. When looking at the Nu time series, we see large bursts of Nu surrounded by ``quiet'' intervals where $\tnu(\text{t}) \approx 1$. During these intervals, vertical motion is absent and the flow largely moves horizontally. This is allowed by the use of lateral periodicity. Zonal flow is observed when either one or both plates have a stress-free boundary condition. Zonal flow is additionally observed for no-slip plates in a special case, i.e. when the sidewall is periodic and $\Gamma < \Gamma_{min}$, and thus the flow has limited freedom. This results in a similar Nu(t) structure. A difference between these cases is that zonal flow is observed \emph{upwards} from a critical Ra, while for the low $\Gamma$ case with lateral periodicity, the bursting behaviour of Nu(t) is only observed for a Ra \emph{lower} than a critical Ra. In all cases, zonal flow is only observed for periodic sidewalls in two-dimensional RB.

We have quantified the effect of the sidewall on Nu for no-slip plates when the flow is roll-like using for no-slip sidewalls, stress-free sidewalls and lateral periodicity. As expected, Nu is higher for stress-free than for no-slip sidewalls over the evaluated Ra range. Surprisingly, in the stress-free case, Nu is higher for lower $\Gamma$, while the opposite happens for no-slip. This can be attributed to the absence of vertical roll states for stress-free sidewalls. In addition, stress-free boundary conditions result in heat flux, when compared to lateral periodicity for both $\Gamma = 1.5$ and $\Gamma = 2$ and for both of these boundary conditions the aspect-ratio dependence of Nu disappears at high Ra.

From analysing Nu($\Gamma$) for lateral periodicity and no-slip plates, we can conclude that below a certain $\Gamma_{\text{min}}$ the flow becomes zonal. Roll-like flow is supressed, resulting in a lower Nu than larger $\Gamma > \Gamma_{\text{min}}$. The type of zonal flow that results depends strongly on $\Gamma$, as zonal flows with two and three shearing layers are observed for $\Gamma = 1$ and $\Gamma = 0.5$, respectively. $\Gamma_{\text{min}}$ decreases for higher Ra, most likely due to the decreased viscous stresses on the thermal plumes. This allows the plumes to require less freedom to move alongside each other in opposite directions, allowing for roll-like flow.

Finally, in the case of asymmetric plate boundary conditions, where one horizontal plate has a no-slip boundary condition and the other plate a stress-free boundary condition, zonal flow can be observed from a critical Ra onwards, as was the case for two stress-free plates. Below this Ra, the flow remains roll-like, although the mean temperature is not 1/2 as the up-down symmetry is broken due to the different velocity boundary conditions. The plate with the stress-free boundary condition has a larger effect on the mean temperature and it therefore deviates from 1/2 towards the temperature of the SF plate. This same effect was also seen previously with stress-free sidewalls, even though the plate boundary conditions were equal. We wish to highlight that even though the effect is the same, they are due to very different mechanisms. The resulting vertical temperature profile is also completely different between the zonal flow and the roll-like flow regimes. In addition, we observed hysteretic behaviour around the transition from roll-like to zonal flow, similar to the case with two stress-free plates, reported in \cite{gol14}.

Choosing the boundary conditions for a simulation can be an involved, but very relevant, consideration, as results have shown here. Based on computational performance, the laterally periodic simulations with stress-free plates would be the first choice; removing most boundary layers. In two-dimensions this can result in zonal flow with very little resemblance to roll-like RB convection, commonly observed in experiments. Zonal flow is avoided by introducing drag through the no-slip plate boundary condition. However, the roll-like convection that corresponds to this set of boundary conditions requires a minimum aspect-ratio, otherwise zonal flow will emerge. Above this, the aspect-ratio dependence, but not the boundary condition dependence, of the heat flux becomes negligible for $\tra \ge 10^{10}$ at least for stress-free and periodic sidewall. Introducing asymmetry in the system, between either the plates or the sidewall and the plates, can result in zonal flow but also surprising effects as unexpected mean temperatures deviating from $1/2$. Most, if not all, of these boundary conditions effects are qualitatively expected to be non-existent or at least much weaker in three-dimensions as compared to two-dimensions. 

\vspace{0.5cm}
\noindent \textbf{Acknowledgements:} The authors thank D. Goluskin, H. Johnston and S. Grossmann for useful discussions. The work was supported by the Foundation for Fundamental Research on Matter (FOM) and the National Computing Facilities (NCF), both sponsored by NWO.

\bibliography{bcpaper.bbl}

\appendix
\section{Details of numerical simulations}

\begin{table*}
\centering
\caption{The columns from left to right indicate Ra, Pr, $\Gamma$, the wall BC, the plate BC, the resolution in horizontal and vertical direction $N_x \times N_z$, the number of grid points in the thermal boundary layer $\#\{n~|~n \in \{1,2,...,N_z\}~\wedge~Z(n) < L/(2\tnu)\}$ (where $Z(n)$ is the vertical grid point distribution), the average Kolmogorov length scale in the flow compared to the largest grid length used somewhere in the grid $\max(\delta_x,\delta_z)/\eta$ with$\eta/L = \tpr^{1/2}/[\tra(\tnu-1)]^{1/4}$, the mean Nu over the full simulation and the averaging time in free fall time units $\tau_f$.}
    \begin{tabular}{| c | c | c | c | c | c | c | c | c | c | c |}
    \hline
    Ra & Pr & $\Gamma$ &  wall BC & plate BC  &$N_x \times N_z$ &  $N_{BL}$ & $\frac{\max(\delta_x,\delta_z)}{\eta}$ &$\tnu$  & $\tau_f$\\
    \hline
    $10^7$            & 1 & 2.00 & PD & AS & $512\times 256$   & 11 & 0.47 & 21.6 & 750\\
    $2.15\times10^7$  & 1 & 2.00 & PD & AS & $512\times 256$   & 10 & 0.59 & 25.8 & 900\\
    $4.64\times10^7$  & 1 & 2.00 & PD & AS & $512\times 256$   & 9 & 0.76 & 32.0 & 900\\
    $10^8$            & 1 & 2.00 & PD & AS & $512\times 256$   & 8 & 0.95 & 37.0 & 900\\
    $2.15\times10^8$  & 1 & 2.00 & PD & AS & $1024\times 512$  & 15 & 0.60 & 43.0 & 900\\
    $4.64\times10^8$  & 1 & 2.00 & PD & AS & $1024\times 512$  & 13 & 0.76 & 52.3 & 1000\\
    $10^9$            & 1 & 2.00 & PD & AS & $3072\times 1024$   & 118 & 0.20 & 9.8 & 10000\\
    $2.15\times10^9$  & 1 & 2.00 & PD & AS & $3072\times 1024$  & 128 & 0.25 & 8.6 & 14000\\
    $10^9$            & 1 & 1.00 & PD & NS & $512\times 512$   & 19 & 0.79 & 27.9 & 1700\\
    $10^9$            & 1 & 1.20 & PD & NS & $512\times 512$   & 18 & 0.96 & 29.1 & 1500\\
    $10^9$            & 1 & 1.25 & PD & NS & $512\times 512$   & 16 & 1.18 & 54.8 & 600\\
    $10^9$            & 1 & 1.30 & PD & NS & $512\times 512$   & 17 & 1.22  & 53.7 & 400\\
    $10^9$            & 1 & 1.40 & PD & NS & $1024\times 512$  & 11 & 0.66  & 53.7 & 1500\\
    $10^9$            & 1 & 1.60 & PD & NS & $1024\times 512$  & 11 & 0.74  & 51.0 & 1500\\
    $10^9$            & 1 & 1.80 & PD & NS & $1024\times 512$  & 12 & 0.82  & 48.5 & 1500\\
    $10^9$            & 1 & 2.00 & PD & NS & $1024\times 512$  & 12 & 0.91  & 48.3 & 1200\\
    $10^9$            & 1 & 2.20 & PD & NS & $1024\times 512$  & 12 & 1.00  & 47.9 & 1200\\
    $10^9$            & 1 & 2.40 & PD & NS & $1600\times 512$  & 12 & 0.70  & 47.6 & 900\\
    $10^9$            & 1 & 2.60 & PD & NS & $1600\times 512$  & 11 & 0.78  & 54.8 & 1000\\
    $10^9$            & 1 & 2.80 & PD & NS & $1600\times 512$  & 11 & 0.84  & 53.4 & 1300\\
    $10^9$            & 1 & 3.00 & PD & NS & $1600\times 512$  & 11 & 0.88  & 50.9 & 1200\\
    $10^{10}$         & 1 & 1.00 & PD & NS & $1024\times 1024$ & 28 & 0.85  & 59.1 & 400\\
    $10^{10}$         & 1 & 1.05 & PD & NS & $1024\times 1024$ & 51 & 0.86  & 49.2 & 500\\
    $10^{10}$         & 1 & 1.10 & PD & NS & $1024\times 1024$ & 32 & 1.08 & 102 & 1100\\
    $10^{10}$         & 1 & 1.20 & PD & NS & $1024\times 1024$ & 18 & 1.17 & 99 & 600\\
    $10^{10}$         & 1 & 1.40 & PD & NS & $1440\times 1024$ & 18 & 0.96  & 95.8 & 600\\
    $10^{10}$         & 1 & 1.60 & PD & NS & $1440\times 1024$ & 18 & 1.10  & 95.7 & 600\\
    $10^{10}$         & 1 & 1.80 & PD & NS & $1440\times 1024$ & 18 & 1.24  & 96.4 & 700\\
    $10^{10}$         & 1 & 2.00 & PD & NS & $1440\times 1024$ & 42 & 1.37  & 94.9 & 400\\
    $10^{10}$         & 1 & 2.20 & PD & NS & $2048\times 1024$ & 19 & 1.05  & 93.7 & 1000\\
    $10^{10}$         & 1 & 2.40 & PD & NS & $2048\times 1024$ & 19 & 1.15  & 95.3 & 1000\\
    $10^{10}$         & 1 & 2.60 & PD & NS & $2048\times 1024$ & 19 & 1.24  & 93.1 & 1100\\
    $10^{10}$         & 1 & 2.80 & PD & NS & $2048\times 1024$ & 19 & 1.34  & 93.3 & 1000\\
    $10^{10}$         & 1 & 3.00 & PD & NS & $2048\times 1024$ & 23 & 1.43  & 92.6 & 1000\\
    $10^{11}$         & 1 & 1.00 & PD & NS & $2048\times 2048$ & 61 & 0.82  & 77.9 & 300\\
    $10^{11}$         & 1 & 1.05 & PD & NS & $2400\times 2048$ & 50 & 0.92 & 198 & 200\\
    $10^{11}$         & 1 & 1.10 & PD & NS & $2400\times 2048$ & 51 & 0.95 & 189 & 200\\
    $10^{11}$         & 1 & 1.40 & PD & NS & $2400\times 2048$ & 29 & 1.23 & 197 & 300\\
    $10^{11}$         & 1 & 1.80 & PD & NS & $2560\times 2048$ & 30 & 1.48 & 196 & 200\\
    $10^{11}$         & 1 & 2.20 & PD & NS & $4096\times 2048$ & 31 & 1.11 & 187 & 200\\
    $10^{11}$         & 1 & 2.60 & PD & NS & $4096\times 2048$ & 31 & 1.32 & 188 & 200\\
    $10^{11}$         & 1 & 3.00 & PD & NS & $4096\times 2048$ & 31 & 1.52 & 186 & 200\\
    $10^8$            & 1 & 1.50 & PD & NS & $512\times 256$   & 15 & 0.66  & 26.1 & 600\\
    $2.15\times10^8$  & 1 & 1.50 & PD & NS & $1024\times 512$  & 24 & 0.42  & 33.1 & 400\\
    $4.46\times10^8$  & 1 & 1.50 & PD & NS & $1024\times 512$  & 20 & 0.54  & 41.1 & 400\\
    $10^9$            & 1 & 1.50 & PD & NS & $1024\times 512$  & 17 & 0.69  & 51.3 & 400\\
    $2.15\times10^9$  & 1 & 1.50 & PD & NS & $2048\times 1024$ & 31 & 0.44  & 63.3 & 400\\
    $4.64\times10^9$  & 1 & 1.50 & PD & NS & $2048\times 1024$ & 27 & 0.56  & 75.5 & 400\\
    $10^{10}$         & 1 & 1.50 & PD & NS & $3072\times 1536$ & 38 & 0.48  & 94.4 & 300\\
    $10^8$            & 1 & 2.00 & PD & NS & $512\times 256$   & 15 & 0.66  & 26.1 & 600\\
    $2.15\times10^8$  & 1 & 2.00 & PD & NS & $512\times 256$   & 9  & 1.11  & 31.2 & 2600\\
    $4.46\times10^8$  & 1 & 2.00 & PD & NS & $512\times 256$   & 8  & 1.42  & 38.9 & 2400\\
    $2.15\times10^9$  & 1 & 2.00 & PD & NS & $1024\times 512$  & 10 & 1.17  & 61.1 & 1100\\
    $4.64\times10^9$  & 1 & 2.00 & PD & NS & $1024\times 512$ & 10 & 1.50  & 76.3 & 1000\\
    \hline
  \end{tabular}
\label{table1}
\end{table*}

\begin{table*}
\centering
\caption{The columns from left to right indicate Ra, Pr, $\Gamma$, the wall BC, the plate BC, the resolution in horizontal and vertical direction $N_x \times N_z$, the number of grid points in the thermal boundary layer $\#\{n~|~n \in \{1,2,...,N_z\}~\wedge~Z(n) < L/(2\tnu)\}$ (where $Z(n)$ is the vertical grid point distribution), the average Kolmogorov length scale in the flow compared to the largest grid length used somewhere in the grid $\max(\delta_x,\delta_z)/\eta$ with $\eta/L = \tpr^{1/2}/[\tra(\tnu-1)]^{1/4}$, the mean Nu over the full simulation and the averaging time in free fall time units $\tau_f$.}
    \begin{tabular}{| c | c | c | c | c | c | c | c | c | c | c |}
    \hline
    Ra & Pr & $\Gamma$ &  wall BC & plate BC  &$N_x \times N_z$ &  $N_{BL}$ & $\frac{\max(\delta_x,\delta_z)}{\eta}$ &$\tnu$      & $\tau_f$\\
    \hline
    $10^8$            & 1 & 1.50 & SF & NS & $512\times 256$   & 13 & 0.68  & 29.1 & 600\\
    $2.15\times10^8$  & 1 & 1.50 & SF & NS & $1024\times 512$  & 23 & 0.4l  & 36.1 & 400\\
    $4.46\times10^8$  & 1 & 1.50 & SF & NS & $1024\times 512$  & 19 & 0.55  & 45.1 & 400\\
    $10^9$            & 1 & 1.50 & SF & NS & $1024\times 512$  & 16 & 0.70  & 56.0 & 400\\
    $2.15\times10^9$  & 1 & 1.50 & SF & NS & $2048\times 1024$ & 30 & 0.45  & 66.1 & 400\\
    $4.64\times10^9$  & 1 & 1.50 & SF & NS & $2048\times 1024$ & 25 & 0.58  & 83.7 & 400\\
    $10^{10}$         & 1 & 1.50 & SF & NS & $3072\times 1536$ & 36 & 0.49 & 101 & 300\\
    $10^8$            & 1 & 2.00 & SF & NS & $512\times 256$   & 9 & 0.90  & 29.7 & 500\\
    $2.15\times10^8$  & 1 & 2.00 & SF & NS & $576\times 320$  & 11 & 1.03  & 36.4 & 600\\
    $4.46\times10^8$  & 1 & 2.00 & SF & NS & $1024\times 512$  & 17 & 0.74  & 45.8 & 600\\
    $10^9$            & 1 & 2.00 & SF & NS & $1024\times 512$  & 15 & 0.95  & 56.8 & 500\\
    $2.15\times10^9$  & 1 & 2.00 & SF & NS & $1152\times 576$ & 15 & 1.07  & 69.0 & 500\\
    $4.64\times10^9$  & 1 & 2.00 & SF & NS & $1280\times 640$ & 15 & 1.22  & 81.9 & 600\\
    $10^{10}$         & 1 & 2.00 & SF & NS & $2048\times 1440$ & 41 & 0.98 & 101 & 500\\
    $10^8$            & 1 & 1.00 & SF & NS & $512\times 512$   & 18 & 0.47  & 33.4 & 400\\
    $2.15\times10^8$  & 1 & 1.00 & SF & NS & $512\times 512$  & 17 & 0.57 & 35.4 & 500\\
    $4.46\times10^8$  & 1 & 1.00 & SF & NS & $512\times 512$  & 19 & 0.74 & 44.5 & 500\\
    $10^9$            & 1 & 1.00 & SF & NS & $512\times 512$   & 11 & 0.94  & 54.6 & 500\\
    $2.15\times10^9$  & 1 & 1.00 & SF & NS & $1024\times 1024$ & 29 & 0.61  & 70.1 & 600\\
    $4.46\times10^9$  & 1 & 1.00 & SF & NS & $1024\times 1024$ & 23 & 0.79  & 92.9 & 500\\
    $10^{10}$         & 1 & 1.00 & SF & NS & $2048\times 1441$ & 35 & 0.49 & 103 & 200\\
    $2.15\times10^{10}$ & 1 & 1.00 & SF & NS & $2048\times 1441$ & 28 & 0.64 & 135 & 200\\
    $4.64\times10^{10}$ & 1 & 1.00 & SF & NS & $2048\times 1441$ & 22 & 1.00 & 180 & 300\\
    $10^{11}$         & 1 & 1.00 & SF & NS & $3240\times 3072$ & 59 & 0.66 &217 & 100\\
    $10^8$            & 1 & 0.33 & SF & NS & $512\times 512$   & 16 & 0.16 & 37.8 & 500\\
    $2.15\times10^8$  & 1 & 0.33 & SF & NS & $512\times 512$  & 13 & 0.20 &  46.5 & 500\\
    $4.46\times10^8$  & 1 & 0.33 & SF & NS & $512\times 512$  & 11 & 0.26 &  57.0 & 500\\
    $10^9$            & 1 & 0.33 & SF & NS & $1024\times 1024$ & 35 & 0.17 & 73.5 & 200\\
    $2.15\times10^9$  & 1 & 0.33 & SF & NS & $640\times 768$ & 18 & 0.35 & 97.8 & 400\\
    $4.46\times10^9$  & 1 & 0.33 & SF & NS & $640\times 768$ & 15 & 0.45 & 125 & 400\\
    $10^{10}$         & 1 & 0.33 & SF & NS & $1024\times 1280$ & 27 & 0.36 & 154 & 500\\
    $2.15\times10^{10}$ & 1 & 0.33 & SF & NS & $1441\times 1600$ & 31 & 0.33 & 199 & 400\\
    $4.64\times10^{10}$ & 1 & 0.33 & SF & NS & $2048\times 3072$ & 66 & 0.29 & 240 & 100\\
    $10^{11}$         & 1 & 0.33 & SF & NS & $3072\times 3456$ & 69 & 0.25 & 280 & 200\\
    $10^8$            & 1 & 0.30 & SF & NS & $512\times 512$   & 17 & 0.15  & 38.5 & 400\\
    $10^8$            & 1 & 0.35 & SF & NS & $512\times 512$   & 17 & 0.17  & 37.3 & 500\\
    $10^8$            & 1 & 0.40 & SF & NS & $512\times 512$   & 18 & 0.19  & 36.1 & 500\\
    $10^8$            & 1 & 0.45 & SF & NS & $512\times 512$   & 18 & 0.21  & 35.0 & 500\\
    $10^8$            & 1 & 0.50 & SF & NS & $512\times 512$   & 19 & 0.23  & 34.0 & 500\\
    $10^8$            & 1 & 0.55 & SF & NS & $512\times 512$   & 19 & 0.26  & 33.1 & 500\\
    $10^8$            & 1 & 0.60 & SF & NS & $512\times 512$   & 20 & 0.28  & 32.0 & 400\\
    $10^8$            & 1 & 0.65 & SF & NS & $512\times 512$   & 20 & 0.30  & 31.5 & 500\\
    $10^8$            & 1 & 0.70 & SF & NS & $512\times 512$   & 21 & 0.32  & 31.0 & 500\\
    $10^8$            & 1 & 0.75 & SF & NS & $512\times 512$   & 21 & 0.34  & 30.5 & 500\\
    $10^8$            & 1 & 0.80 & SF & NS & $512\times 512$   & 21 & 0.36  & 29.6 & 400\\
    $10^8$            & 1 & 0.85 & SF & NS & $512\times 512$   & 21 & 0.38  & 29.5 & 500\\
    $10^8$            & 1 & 0.90 & SF & NS & $512\times 512$   & 22 & 0.40  & 28.5 & 400\\
    $10^8$            & 1 & 0.95 & SF & NS & $512\times 512$   & 22 & 0.43  & 28.8 & 500\\
    \hline
  \end{tabular}
\label{table2}
\end{table*}

\begin{table*}
\centering
\caption{The columns from left to right indicate Ra, Pr, $\Gamma$, the wall BC, the plate BC, the resolution in horizontal and vertical direction $N_x \times N_z$, the number of grid points in the thermal boundary layer $\#\{n~|~n \in \{1,2,...,N_z\}~\wedge~Z(n) < L/(2\tnu)\}$ (where $Z(n)$ is the vertical grid point distribution), the average Kolmogorov length scale in the flow compared to the largest grid length used somewhere in the grid $\max(\delta_x,\delta_z)/\eta$ with $\eta/L = \tpr^{1/2}/[\tra(\tnu-1)]^{1/4}$, the mean Nu over the full simulation and the averaging time in free fall time units $\tau_f$.}
    \begin{tabular}{| c | c | c | c | c | c | c | c | c | c | c |}
    \hline
    Ra & Pr & $\Gamma$ &  wall BC & plate BC  &$N_x \times N_z$ &  $N_{BL}$ & $\frac{\max(\delta_x,\delta_z)}{\eta}$ &$\tnu$      & $\tau_f$\\
    \hline
    $10^8$            & 1 & 1.00 & NS & NS & $256\times 256$   & 10 & 1.12 & 26.1 & 300\\
    $2.15\times10^8$  & 1 & 1.00 & NS & NS & $512\times 512$  & 20 & 0.76 & 32.6 & 300\\
    $4.46\times10^8$  & 1 & 1.00 & NS & NS & $512\times 512$  & 18 & 1.00 & 40.8 & 400\\
    $10^9$            & 1 & 1.00 & NS & NS & $512\times 512$   & 19 & 1.33 & 50.5 & 800\\
    $2.15\times10^9$  & 1 & 1.00 & NS & NS & $1024\times 1440$ & 52 & 0.85 & 61.5 & 400\\
    $4.46\times10^9$  & 1 & 1.00 & NS & NS & $1024\times 1440$ & 44 & 1.08 & 75.3 & 400\\
    $10^{10}$         & 1 & 1.00 & NS & NS & $1024\times 1440$ & 37 & 1.45 & 98.0 & 200\\
    $2.15\times10^{10}$ & 1 & 1.00 & NS & NS & $2048\times 2048$ & 55 & 0.94 & 125 & 200\\
    $4.64\times10^{10}$ & 1 & 1.00 & NS & NS & $3072\times 3072$ & 85 & 0.80 & 145 & 100\\
    $10^{11}$         & 1 & 1.00 & NS & NS & $3072\times 3072$ & 72 & 1.04 & 187 & 200\\
    $10^8$            & 1 & 0.33 & NS & NS & $256\times 256$   & 15 & 0.36 & 19.2 & 1000\\
    $2.15\times10^8$  & 1 & 0.33 & NS & NS & $512\times 512$  & 13 & 0.34 & 23.2 & 1000\\
    $4.46\times10^8$  & 1 & 0.33 & NS & NS & $512\times 512$  & 14 & 0.38 & 29.8 & 900\\
    $10^9$            & 1 & 0.33 & NS & NS & $512\times 512$ & 24 & 0.76 & 36.6 & 300\\
    $2.15\times10^9$  & 1 & 0.33 & NS & NS & $512\times 576$ & 17 & 0.49 & 46.3 & 200\\
    $4.46\times10^9$  & 1 & 0.33 & NS & NS & $512\times 1024$ & 33 & 0.63 & 59.3 & 300\\
    $10^{10}$         & 1 & 0.33 & NS & NS & $768\times 1440$ & 43 & 0.59 & 87.9 & 200\\
    $2.15\times10^{10}$ & 1 & 0.33 & NS & NS & $768\times 1440$ & 34 & 0.77 & 120 & 300\\
    $4.64\times10^{10}$ & 1 & 0.33 & NS & NS & $2048\times 2048$ & 26 & 1.01 & 172 & 300\\
    $10^{11}$         & 1 & 0.33 & NS & NS & $2048\times 3072$ & 63 & 0.51 & 200 & 100\\
    $10^{7}$         & 40 & 2.00 & PD & SF & $2048\times 1024$ & 41 & 0.02 & 37.4 & 200\\
    $10^{8}$         & 40 & 2.00 & PD & SF & $3072\times 2048$ & 56 & 0.03 & 87.9 & 200\\
    $10^{9}$         & 40 & 2.00 & PD & SF & $4096\times 2048$ & 40 & 0.05 & 168 & 100\\
    $10^{6}$         & 1 & 0.50 & PD & NS & $256\times 256$ & 65 & 0.07 & 2.61 & 90000\\
    $10^{7}$         & 1 & 0.50 & PD & NS & $512\times 512$ & 113 & 0.07 & 3.41 & 300000\\
    $10^{8}$         & 1 & 0.50 & PD & NS & $512\times 512$ & 96 & 0.13 & 4.32 & 200000\\
    $10^{9}$         & 1 & 0.50 & PD & NS & $1024\times 1024$ & 171 & 0.13 & 5.79 & 20000\\
    $10^{11}$         & 1 & 0.50 & PD & NS & $2048\times 3072$ & 105 & 0.43 & 95.4 & 200\\
    $10^{12}$         & 1 & 0.50 & PD & NS & $3072\times 4096$ & 73 & 0.67 & 282 & 100\\
    \hline
  \end{tabular}
\label{table3}
\end{table*}
\end{document}